\documentclass[prb,aps,twocolumn,showpacs,citeautoscript,superscriptaddress,amsmath]{revtex4-1}
\usepackage{graphicx}% Include figure files \usepackage{bm}% bold math
\usepackage{amssymb}

\newcommand{\eq}{\begin{equation}} \newcommand{\eqn}{\end{equation}}

\usepackage{color}  \def\unit#1{\ \mathrm{#1}}

\definecolor{OrchidDark}{rgb}{0.6000,0.1961,0.8000}
\usepackage{color}  \def\unit#1{\ \mathrm{#1}}

\begin{document}

\title{From laterally modulated two-dimensional electron gas towards
  artificial graphene}

\author{L.~N\'{a}dvorn\'{i}k} \email{nadvl@fzu.cz}
\affiliation{Faculty of Mathematics and Physics, 
Charles University, Ke Karlovu 3, 121 16 Praha 2, Czech Republic\\}
\affiliation{Institute of Physics, ASCR, v.v.i., 
Cukrovarnick\'{a} 10, 162 53 Praha 6, Czech Republic\\}
\author{M.~Orlita}
\affiliation{Laboratoire
National des Champs Magn\'etiques Intenses, CNRS-UJF-UPS-INSA, 25, avenue des
Martyrs, 38042 Grenoble, France\\}
\affiliation{Institute of Physics, ASCR, v.v.i., 
Cukrovarnick\'{a} 10, 162 53 Praha 6, Czech Republic\\}
\affiliation{Faculty of Mathematics and Physics, 
Charles University, Ke Karlovu 3, 121 16 Praha 2, Czech Republic\\}
\author{N.~A.~Goncharuk}
\affiliation{Institute of Physics, ASCR, v.v.i., 
Cukrovarnick\'{a} 10, 162 53 Praha 6, Czech Republic\\}
\author{L.~\ Smr\v{c}ka}
\affiliation{Institute of Physics, ASCR, v.v.i., 
Cukrovarnick\'{a} 10, 162 53 Praha 6, Czech Republic\\}
\author{V.~Nov\'{a}k}
\affiliation{Institute of Physics, ASCR, v.v.i., 
Cukrovarnick\'{a} 10, 162 53 Praha 6, Czech Republic\\}
\author{V.~Jurka}
\affiliation{Institute of Physics, ASCR, v.v.i., 
Cukrovarnick\'{a} 10, 162 53 Praha 6, Czech Republic\\}
\author{K.~Hru\v{s}ka}
\affiliation{Institute of Physics, ASCR, v.v.i., 
Cukrovarnick\'{a} 10, 162 53 Praha 6, Czech Republic\\}
\author{Z.~V\'{y}born\'y}
\affiliation{Institute of Physics, ASCR, v.v.i., 
Cukrovarnick\'{a} 10, 162 53 Praha 6, Czech Republic\\}
\author{Z. R. Wasilewski}
\affiliation{Institute for Microstructural Sciences, NRC, 
Ottawa, Ontario, Canada K1A~0R6}
\author{M.~Potemski}
\affiliation{Laboratoire
National des Champs Magn\'etiques Intenses, CNRS-UJF-UPS-INSA, 25, avenue des
Martyrs, 38042 Grenoble, France\\}
%\affiliation{Laboratoire
%National des Champs Magn\'etiques Intenses, CNRS-UJF-UPS-INSA, 25, avenue des
%Martyrs, 38042 Grenoble, France}
\author{K.~V\'{y}born\'y}
\affiliation{Institute of Physics, ASCR, v.v.i., 
Cukrovarnick\'{a} 10, 162 53 Praha 6, Czech Republic\\}
\affiliation{Department of Physics, University at Buffalo--SUNY,
Buffalo, New York 14260, USA}

\date{November 27, 2011}


% 73.22.Pr 	Electronic structure of graphene
% 73.21.Cd      Superlattices (in sect. Electron states and collective
%                excitations in multilayers, quantum wells, mesoscopic,
%                and nanoscale systems
% 78.67.Pt      Multilayers; superlattices; photonic structures;
%               metamaterials (Optical properties of low-dimensional,
%               mesoscopic, and nanoscale materials and structures)

\begin{abstract}

Cyclotron resonance has been measured in far-infrared transmission of
GaAs/Al$_x$Ga$_{1-x}$As heterostructures with an etched hexagonal
lateral superlattice. Non-linear dependence of the resonance position on
magnetic field was observed as well as its splitting into several
modes. Our explanation, based on a perturbative calculation, describes
the observed phenomena as a weak effect of the lateral potential on
the two-dimensional electron gas.  Using this approach, we found a
correlation between parameters of the lateral patterning and the
created effective potential and obtain thus insights on how the
electronic miniband structure has been tuned. The miniband dispersion
was calculated using a simplified model and allowed us to formulate
four basic criteria that have to be satisfied to reach graphene-like
physics in such systems.

\end{abstract}

\maketitle

\section{Introduction}
\label{Intro}

\begin{figure} [b]
	\begin{center}
	\includegraphics[width=0.9\hsize,
	angle=0]{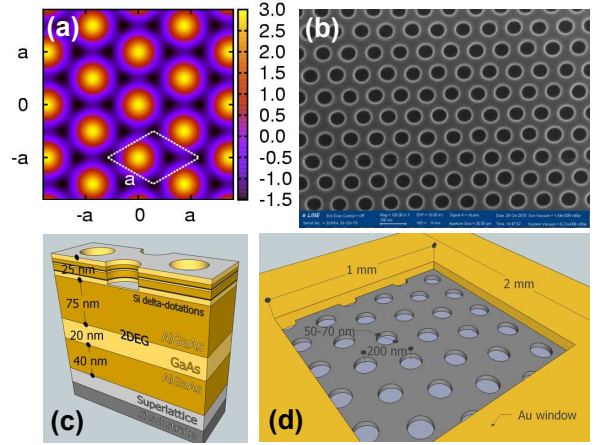}
	\caption{\small{(color online) Artificial graphene: Part (a):
	hexagonal potential of Eq.~(\ref{eq-02}) defining the AG
        (in units of $V_0$). Dark regions
	correspond to carbon atoms in the real graphene lattice;
        the primitive cell is indicated. (b): Scanning electron
        microscope image of the surface of
	one of the samples, (c): its atomic layer structure (2DEG
        depth $\approx 100$~nm) and (d): the sample layout (not to scale).}}
	\label{img:1}
	\end{center}
\end{figure}

The range of approaches to explore Dirac fermions in condensed-matter
physics has recently been extended beyond natural
graphene\cite{NovoselovNature05,ZhangNature05,GeimNatureMaterial07} to
artificially created lattices whose properties such as inter-site
coupling or lattice constant can be tuned.  One approach here is to
subject gas of ultracold atoms to a honeycomb optical
lattice\cite{GrynbergPRL93} giving rise to Dirac cones in dispersion
relations.\cite{ZhuPRL07,WunchNJP08} The same has been proposed for
lithographically patterned two-dimensional electron gases (2DEGs) in
semiconductor heterostructures\cite{ParkPRL08,ParkNL09,GibertiniPRB09}
as shown in Fig.~\ref{img:1}. Apart from
studying the Dirac fermions on their own, an appealing perspective in
such semiconductor-based systems would be to fabricate various
proof-of-principle electronic devices originally proposed for natural
graphene,\cite{DuboisEPJB09} for instance filters and
valves,\cite{RycerzNaturePhys07} Veselago
lenses,\cite{CheianovScience07} or
splitters\cite{Garcia-PomarPRL08}. The electron-beam lithography used
to define the artificial honeycomb crystal potential, dubbed
artificial graphene (AG),\cite{GibertiniPRB09} allows for much easier
control over the device details such as the edge geometry, additional
or missing ``atoms'' than what would require an atom-by-atom
manipulation\cite{Eigler:1990_a} in the case of natural graphene.

The subject of this paper, fabrication and theoretical description of
AG, represents a special case of lateral semiconductor superlattices
(SLs) intensively studied especially in the nineties. The body of
widely explored phenomena can be roughly divided into two classes:
classical and quantum-mechanical. Magneto-plasmons in far-infrared
transmission\cite{KernPRL91} and commensurability
oscillations\cite{WeissEPL89} in a modulated 2DEG can be largely
explained without invoking quantum
mechanics.\cite{MikhailovPRB96,BeenakkerPRL89} On the other hand, SLs
under strong magnetic fields, which leads to the magnetic
breakdown,\cite{StredaPRB90,Gvozdikov:2007_a} as well as other more complex
systems\cite{AlbrechtPRL99,OlszewskiPSSB04,Gvozdikov:2007_b} require 
quantum-mechanical ingredients (Bohr-Sommerfeld quantization and
tunneling between semiclassical orbits).

In this context, the Dirac cone in the AG spectrum is a delicate
quantum-mechanical feature just as the Hofstadter
butterfly\cite{HofstadterPRB76} --- Landau bands emerging when
$p$ quanta of magnetic flux pierce $q$ elementary cells of a square
SL.\cite{GeislerPRL04,Pfannkuche:1992_a} Despite promising recent
progress,\cite{GibertiniPRB09,DeSimoniAPL10,SinghaScience11} a
clear-cut evidence of Dirac fermions in AG has not yet been found.
Most likely, such an evidence can be provided in a magneto-transport or
magneto-optical experiment by observing half-integer quantum Hall
effect,\cite{NovoselovNature05,ZhangNature05} or
the unique $\sqrt{B}$ scaling in the optical response due to
inter-Landau level (inter-LL)
transitions.\cite{SadowskiPRL06,JiangPRL07,DeaconPRB07} More careful
engineering of AG structures is likely needed to realize these
experiments.

In this paper, we explore the concept of artificial graphene both
experimentally and theoretically. We use a simplified model of AG to
formulate four straightforward criteria that must be fulfilled to achieve
graphene-like physics in conventional 2D semiconductor
heterostructures. Among these criteria, we focus on formation of a
suitable miniband structure, explain how it is related to the
modulation potential amplitude and show how this quantity can be
measured in far-infrared magnetotransmission. Although the Dirac
fermion physics in modulated semiconductor heterostructures has not
been detected so far (including the experiments presented here), we
conclude that simultaneous fulfilment of all the four criteria should
be technologically feasible.

\section{Theory}
\label{theory}

To arrive at transparent conditions necessary for the realization of
Dirac fermions in the SL miniband structure, we use a
maximally simplified AG model which is effectively
single-parametric.\cite{note1}
The modulation potential $V(\vec{r})$, $\vec{r}=(x,y)$ shown at
Fig.~\ref{img:1}(a) is taken as a sum of three cosine functions
\eq V\left(\vec{r}\right) =
V_0(\cos\vec{g_1}\vec{r}+\cos\vec{g_2}\vec{r}+\cos\vec{g_3}\vec{r}),
\label{eq-02}
\eqn
where $\vec{g_1}=2\pi/a(1/\sqrt{3},1)$,
$\vec{g_2}=2\pi/a(2/\sqrt{3},0)$, $\vec{g_3}=\vec{g_2}-\vec{g_1}$ are
the basis vectors in reciprocal space, $a$ is the
distance between two maxima of $V(\vec{r})$ and $V_0$ is the potential
amplitude. Let us note that $V_0$ has to be positive to obtain
a honeycomb structure rather than a trigonal one.  The 
Hamiltonian $\hat{p}^2/2m^*+V(\vec{r})$ in basis $\mathcal{B}$ of plane waves,
\eq
\mathcal{B}=\Big\{e^{i(\vec{k}+\vec{K}_{n_1n_2})\vec{r}},\
\vec{K}_{n_1n_2}=n_1\vec{g}_1+n_2\vec{g}_2\Big\}, \eqn
is a matrix whose diagonal and
off-diagonal matrix elements stand in a ratio determined by $V_0$, $a$
and the electron effective mass $m^\ast$ (in GaAs, $0.067$ of
the electron vacuum mass $m_0$; $\hat{p}$ is the 2D momentum
operator). Except
for an overall scaling, eigenvalues of the matrix depend on a single
dimensionless parameter
\eq \zeta = \frac{m^{\ast}}{(2\pi\hbar)^2}V_0a^2. \eqn
This parameter is, up to a factor of the order of unity, equal to the
ratio between $V_0$ and the kinetic energy $E_0$ of a free electron ($V_0=0$)
at the $K$-point of the Brillouin zone.

\begin{figure} %[htbp!]
	\begin{center}
	\includegraphics[width=0.34\hsize,
	angle=270]{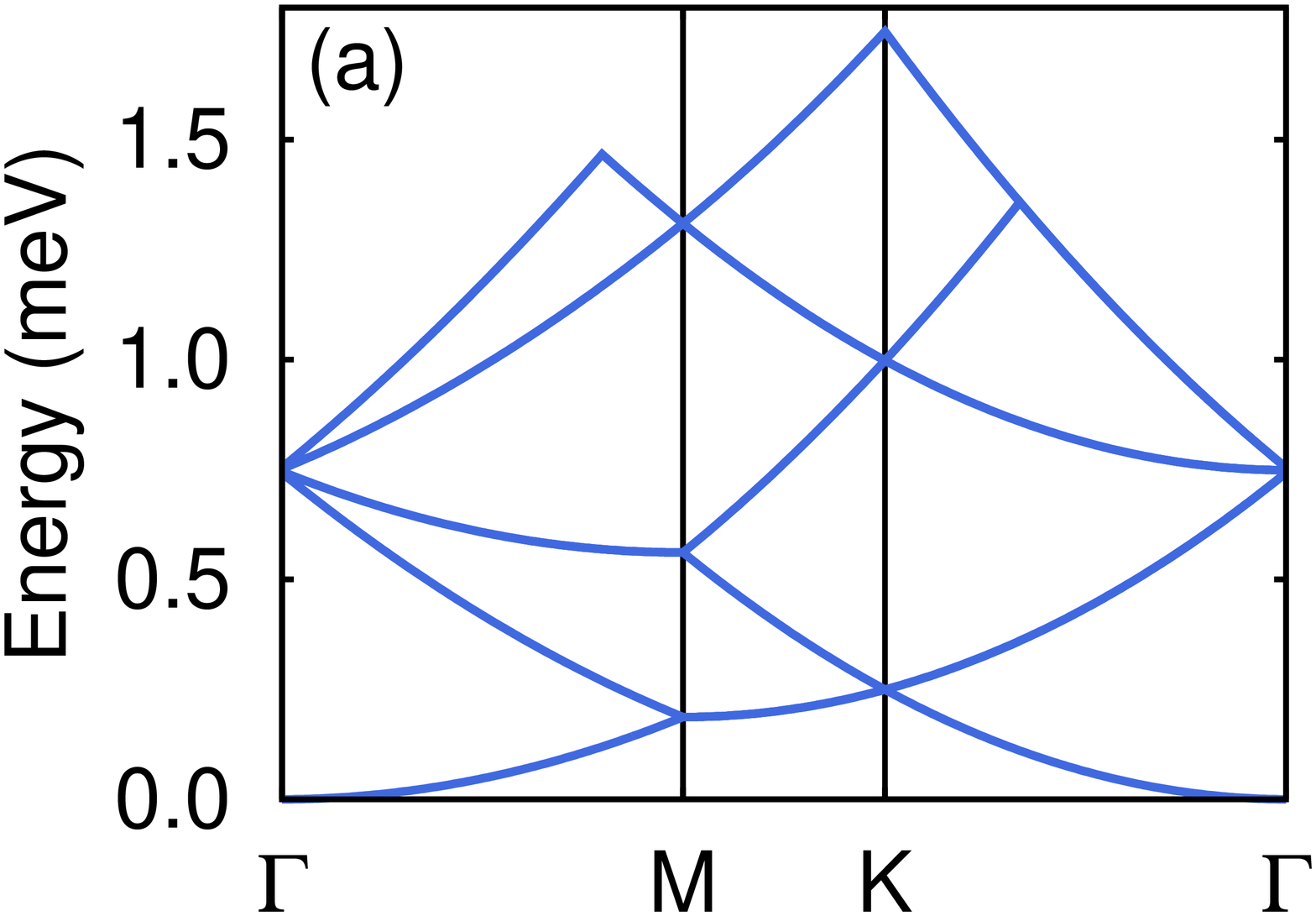}
	\includegraphics[width=0.34\hsize,
	angle=270]{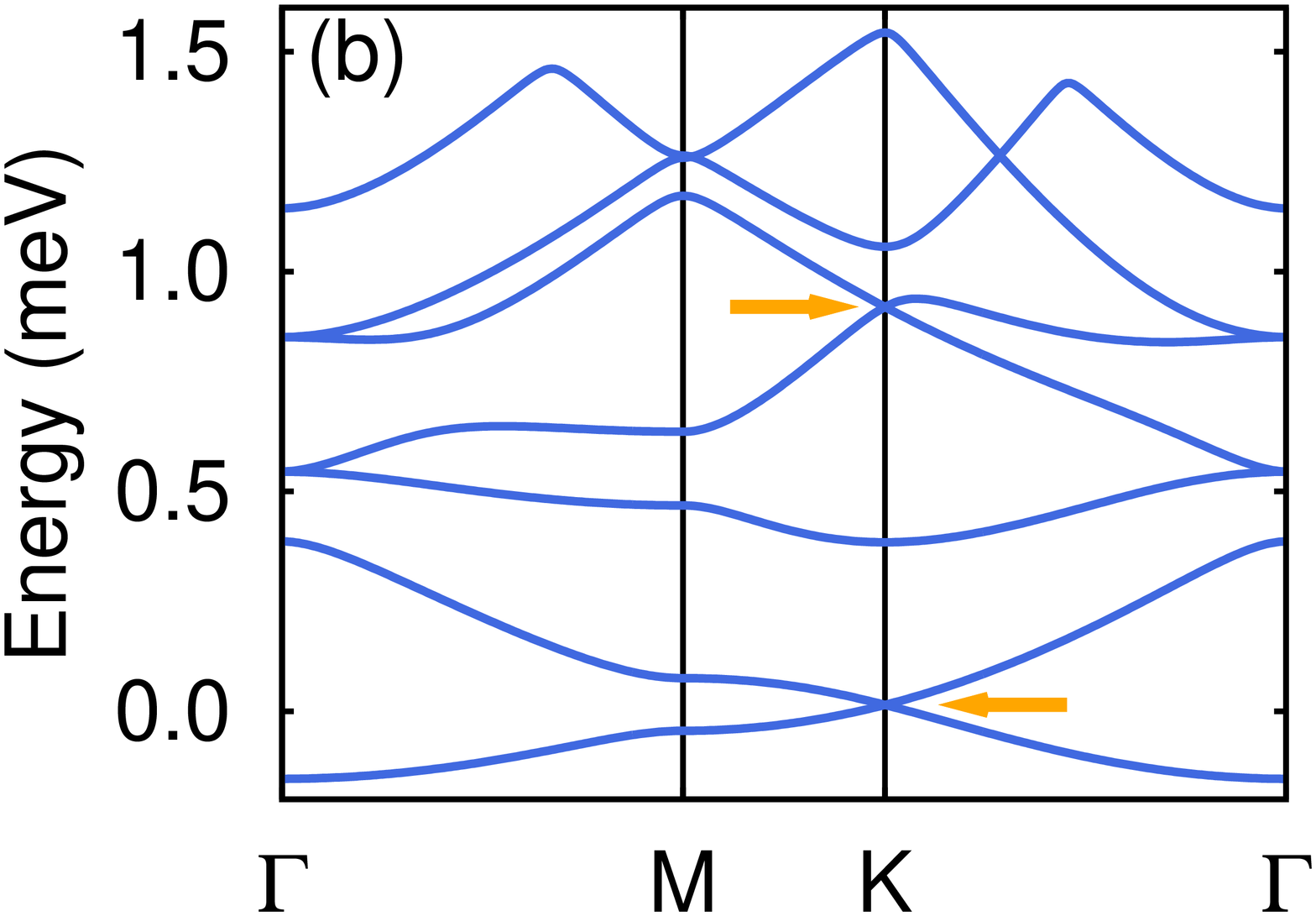}\\
	\includegraphics[width=0.34\hsize,
	angle=270]{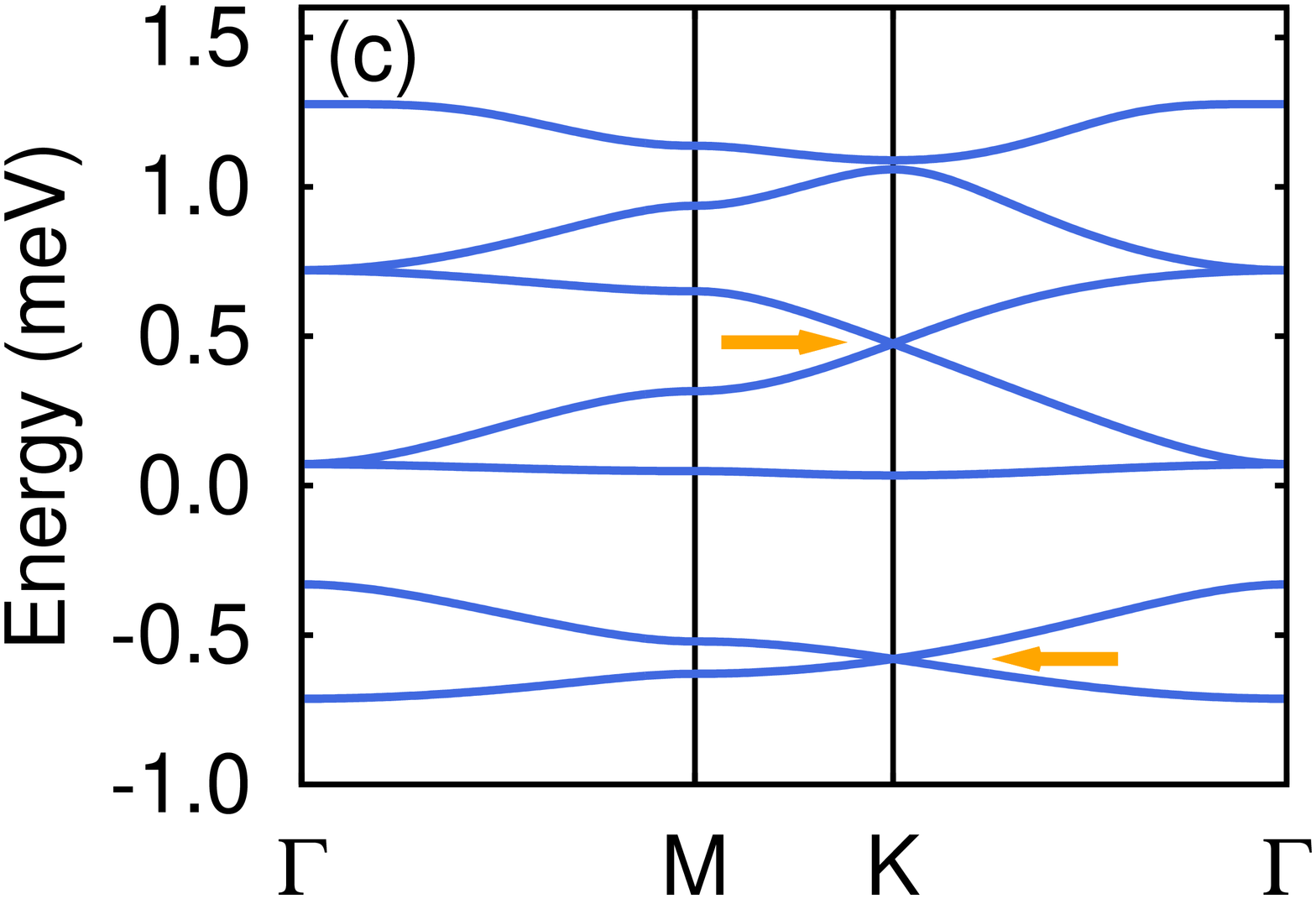}
%	\includegraphics[width=0.34\hsize,
%	angle=270]{pics/150p0-200nm.ps}
	\includegraphics[width=0.34\hsize,
        angle=-90]{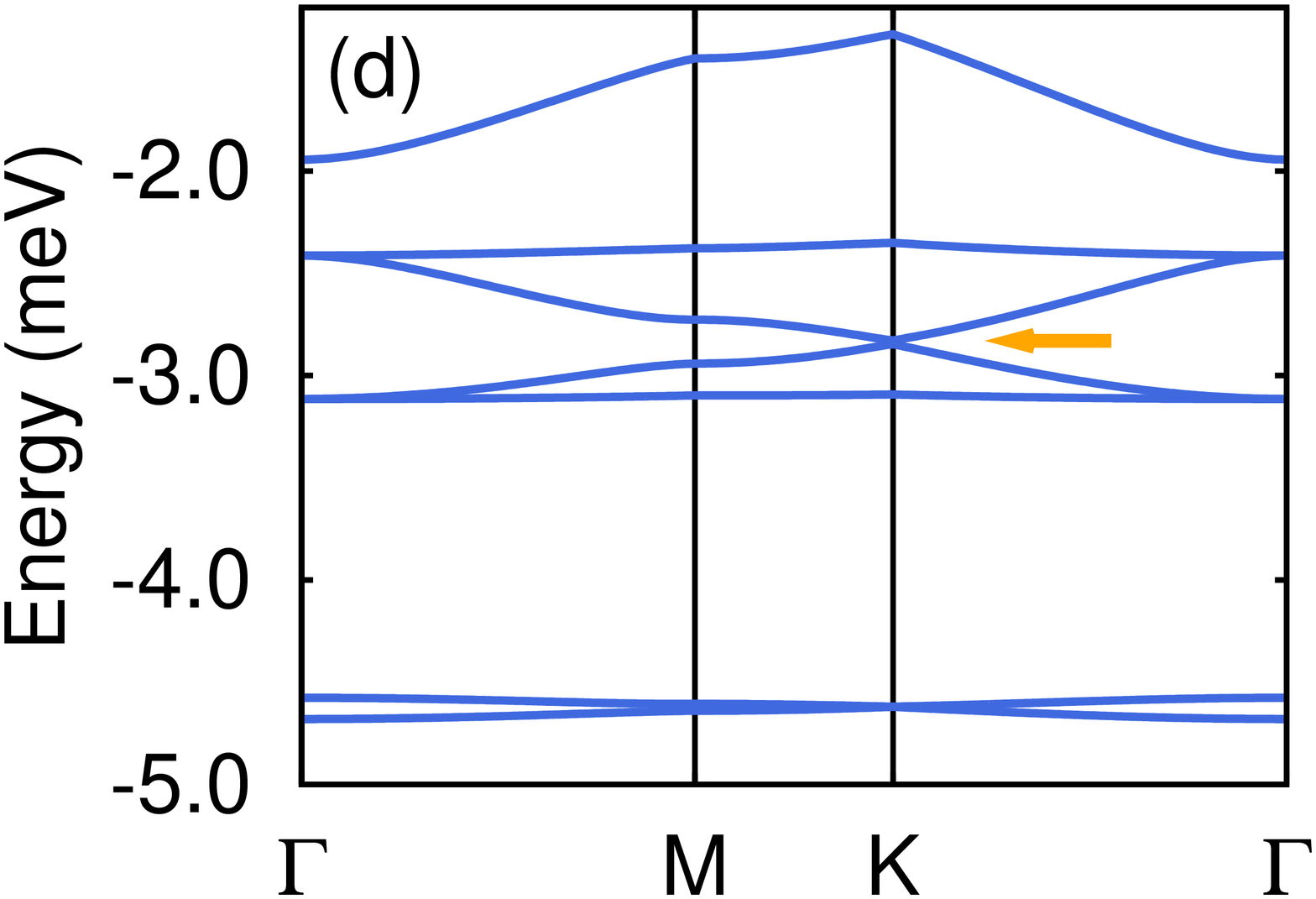}
	\end{center}
    \caption{(Color online) Minibands generated for several values of the
    parameter $\zeta$.  (a): $\zeta=0$, dispersion of a free 2DEG.
    (b): $\zeta=0.3$, lower Dirac cone develops (indicated by arrow),
    the upper one appears but remains covered by other bands.
    (c): $\zeta=0.9$, both Dirac cones fully developed.
    (d): $\zeta=4.0$, tight-binding type narrow minibands form and Dirac cones
    gradually flatten, ultimately becoming again unobservable.
    Energy axis corresponds to GaAs effective mass and
    $a=200\unit{nm}$; panels a, b, c, and d then correspond 
    to $V_0= 0.0$, $0.4$,
    $1.0$ and $4.5\unit{meV}$, respectively.}  \label{img:2}
\end{figure}

Depending on $\zeta$, we obtain miniband spectra that continuously
vary from the free 2DEG, through nearly-free and more
tight-binding-like models, up to nearly flat bands that correspond to
practically isolated (artificial) atoms.  This is illustrated in
Fig.~\ref{img:2}, where we plot miniband structure for four different
values of $\zeta$. In Fig.~\ref{img:2}a, we plot the parabolic
dispersion of a free electron ($\zeta=0$) folded into the newly
created Brillouin zone and then follow the evolution of the miniband
structure with increasing $\zeta$, namely for $\zeta=0.3$, 0.9 and 4.
The present Dirac cones are marked by vertical arrows and their
pseudorelativistic character has been confirmed by analysing the corresponding
wavefunctions.\cite{note2}

Importantly, more than one Dirac cone appears within the seven lowest
lying minibands shown in Fig.~\ref{img:2}. This fact, to the best
of our knowledge so far not mentioned in literature, may significantly
simplify the quest for pseudorelativistic physics as we discuss below.
The Fermi velocity
corresponding to the lower and upper Dirac cone for the modulation
strength $\zeta=0.9$ (Fig.~\ref{img:2}c) is roughly $1.4\times10^4$ and
$3.6\times10^4$~m.s$^{-1}$, respectively.  The expected Fermi velocity
in AG is thus more than order of magnitude lower as compared to
natural graphene, where values around $10^6$~m.s$^{-1}$ are
reported.\cite{NovoselovNature05} Another important characteristics of
the created Dirac cones is their width in energy $E_{DC}$.  A closer
inspection of Fig.~\ref{img:2} reveals that we always get the $E_{DC}$
smaller than $E_0$ and also that $E_{DC}$ strongly depends on
$\zeta$. The typical width of Dirac cones, $E_{DC}$,
is dominantly given by two factors: the size of the Brillouin zone
(i.e. the lattice constant $a$) and the effective mass of the employed
semiconductor system (which defines the kinetic energy $E_0$).

Now we formulate four simple criteria which have to be fulfilled to
achieve Dirac-like AG physics in hexagonally patterned heterostructures.
Unless stated otherwise, we always consider
lattice period $a=200$~nm and the effective mass of electrons in GaAs
$m=0.067m_0$, which match to the samples explored experimentally
below.

(i) \emph{Suitable miniband structure:} The effective strength of the
modulation $\zeta$ has to be tuned to get well-separated and
well-developed Dirac cones. Consistently with our calculations in
Fig.~\ref{img:2}, the range $0.5<\zeta<4.0$
ensures that the Dirac cones do not overlap with other minibands,
as is the case of low $\zeta$ in Fig.~\ref{img:2}b, and also the
cones are not significantly flattened, which gradually happens for
$\zeta>4$ towards the limit of isolated ``atoms''.
If we strictly limit ourselves to the upper Dirac cone, the
effective strengths $\zeta$ somewhat exceeding 4 are still
acceptable. Let us note that the chosen range of $\zeta$ corresponds
to the potential modulation $V_0$ that ranges from 0.6 to
$4.5\unit{meV}$.

(ii) \emph{Fermi level positioning and/or carrier density:} The Fermi
level has to be located in the vicinity of a developed Dirac cone. If
it is well separated from other minibands as required by the previous
point, the corresponding carrier density is easy to estimate.  The
number of states per unit area in one miniband is equal to
$4/(\sqrt{3}a^2)$ with the spin degeneracy included.  The carrier
density then reaches $n \approx 6\times10^{9}$ and $2.5\times
10^{10}$~cm$^{-2}$ with the Fermi level located at the Dirac point of
the lower and upper cone, respectively. If experiments at extremely
low densities (below $10^{10}$~cm$^{-2}$) are to be avoided, we should
preferentially focus on the upper Dirac point. Another way, if
technologically feasible, is to reduce the lattice constant. For
instance, $a=100$~nm implies a very reasonable carrier density of $1.0\times
10^{11}\unit{cm^{-2}}$ for the upper cone. As explained above, the
miniband structure for such reduced lattice constant remains unchanged
(after rescaling the energy axes in Fig.~\ref{img:2} by a factor of
four), provided $V_0$ is increased to keep $\zeta$ constant.

(iii) \emph{Low disorder:} The idealized miniband structure, as
presented in Fig.~\ref{img:2}, is in reality smeared out by
disorder. The minimal requirement is to have the electron mean free
path $l_e=\hbar \mu\sqrt{2\pi n/e^2}$ significantly exceeding
the lattice constant $a$. For 2DEG with density of
$n=10^{11}\unit{cm^{-2}}$ and the relatively low mobility
$\mu=10^5\unit{cm^2/(V.s)}$, the mean free path reaches $l_e\approx
500\unit{nm}$ and still remains well above the technologically
achievable $a$. We emphasize that this is a necessary not a sufficient
condition. 

(iv) \emph{Careful probing:} A clear evidence for the presence of
massless particles will likely come from experiments performed in
magnetic fields, using transport\cite{NovoselovNature05,ZhangNature05}
or optical\cite{SadowskiPRL06,JiangPRL07,DeaconPRB07} methods. The
characteristic spacing of LLs, given by the applied magnetic field $B$
and by the effective Fermi velocity in AG, then should not exceed the
Dirac cone width $E_{DC}$. This condition turns out to be numerically
close to the requirement that the spacing of LLs in an unpatterned
2DEG, $\hbar\omega_c=\hbar eB/m^\ast \approx 2\times
B[\mathrm{T}]$~meV, be small in comparison to the modulation potential
$V_0$ and the width of the particular Dirac cone in AG,
$\hbar\omega_c\ll E_{DC}$. Tolerable magnetic fields are thus hundreds
of millitesla, since $E_{DC}$ reaches about one meV at most in
technologically achievable structures. Such a low magnetic field
requires high quality 2DEG samples.  If we express this quality in
terms of mobility, $\mu B \gtrsim 1$.  It is also important to keep
temperatures low, $kT\ll \hbar\omega_c$, which implicates experiments
in sub-kelvin range for realistic AG structures.

\section{Experiment}
\label{exp}

The lateral modulation of 2DEG is in most cases achieved either by
gating using a specifically patterned metallic
layer\cite{SoibelSST96,HuggerAPL08} or by etching the sample surface
using methods with high spatial
resolution.\cite{KernPRL91,TakaharaJJAP95,GeislerPRL04,%
GeislerPRB05,GibertiniPRB09}
In the more common latter case, arrays of dots or antidots are
fabricated.\cite{HeitmannPT93}  At a given pitch between
dots/antidots, the antidot design allows us to get a
factor of $\sqrt3$ lower lattice constant of AG as compared to array of
dots. The etching depth serves as a parameter tuning strength of the
lateral potential. An especially strong modulation can be achieved by
etching through the 2DEG layer. Such structures, with regions fully
depleted from electrons,\cite{MikhailovPRB96} have been mostly used to
study magneto-plasmon effects\cite{KernPRL91}.  On the other hand,
quantum effects due to miniband structure are typically studied in
shallow-etched samples\cite{GeislerPRL04} such as are subject of our study.

The studied samples have been prepared by etching a shallow array of
holes (i.e. antidots) with a triangular symmetry. The electron beam
lithography and dry etching process (Ar$^{+}$+ SiCl$_4$) have been
employed. We prepared three samples denoted as A, B and C, see
Table~\ref{tab-01}, with the etching depth of 15-25, 20 and 48~nm,
respectively. The diameter of holes was always $\approx60$~nm and
hole-to-hole distance, i.e., our AG lattice constant,
$a=200$~nm. Samples A and B have been prepared from a wafer with 2DEG
in a 20~nm-wide quantum well embedded between Al$_{0.33}$Ga$_{0.67}$As
barriers and located 100~nm below the surface. The electrons in the
well are provided by two $\delta$-doped Si layers, 15~nm
($3\times10^{12}$ cm$^{-2}$) and 25~nm ($2\times10^{12}$ cm$^{-2}$)
deep, see Fig.~\ref{img:1}(c).  The sample C was fabricated from a
simple GaAs/Al$_{0.33}$Ga$_{0.67}$As heterojunction located 115~nm
below the sample surface. The triangular well formed at the interface
was filled by electrons from a Si-doped Al$_{0.33}$Ga$_{0.67}$As
region ($1.5\times10^{18}$ cm$^{-3}$) separated from the interface by
a spacer 25~nm wide.  At all three samples, the lithographically
patterned area ($1\times1$~mm$^2$) was surrounded by a gold frame
(50~nm thick) to define the optically active part for the transmission
experiment, see Fig.~\ref{img:1}(d).

The prepared samples have been studied using the infrared
magneto-spectroscopy technique, in this case having a form of the
Landau level spectroscopy. To measure the magneto-transmission of the
sample, the radiation of globar or mercury lamp was modulated by a
Fourier transform spectrometer. 
We worked with photon energies down to 4~meV at a resolution of
0.125~meV. The radiation was delivered via light-pipe optics to the
sample kept at 2~K inside a superconducting coil and detected by a Si
bolometer, placed directly below the sample. All measurements were
performed in the Faraday configuration with the magnetic field applied
perpendicular to the sample layer.

\begin{figure}[htbp!]
	\begin{center}
	\includegraphics[width=0.7\hsize]{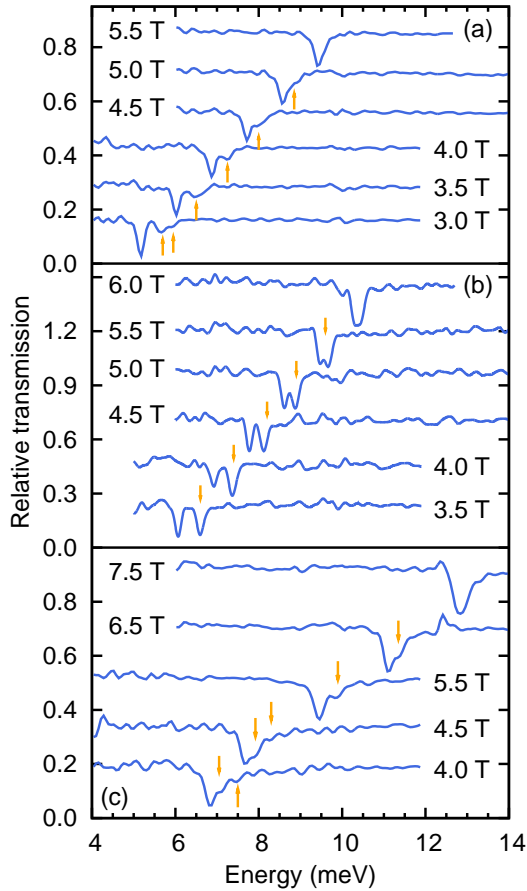}
	\caption{\small{(color online) Relative magneto-transmission spectra
    of studied samples A ($d_{\rm{holes}}\approx15-25$~nm,
    $d_{\rm{2DEG}}=100$~nm), B ($d_{\rm{holes}}=20$~nm,
    $d_{\rm{2DEG}}=100$ nm) and C ($d_{\rm{holes}}=48$~nm,
    $d_{\rm{2DEG}}=115$~nm) in panels (a), (b) and (c),
    respectively. The multi-mode character of cyclotron resonance
    absorption vanishes above
    $B\approx 5, 6$ and 7~T in the sample A, B and C,
    respectively. All spectra are shifted vertically for
    clarity.}}  \label{img:3}
	\end{center}
\end{figure}

Characteristic results of the magneto-transmission experiment are presented in
Fig.~\ref{img:3}.  Relative magneto-transmission spectra are shown, 
i.e. for each photon energy, the transmission at a given field normalized
to the same at $B=0$~T.  A well-defined CR absorption of a
nearly Lorentzian shape is obtained for all three samples at higher
magnetic fields. When the magnetic field is lowered, we observe a more complex
behavior with the CR absorption split into two or even more modes.
The energy distance between these
modes is clearly different in various samples and it is
correlated with the depth of the etched holes, i.e. with the strength
of the modulation potential induced by the lateral patterning. An
unpatterned reference sample, taken from the wafer used for
fabrication of the samples A and B, has been also tested. As
expected, it showed a typical Lorentzian-shaped CR absorption at
the energy of $\hbar\omega_c$ in the whole available range of magnetic
fields.

Since the initial carrier concentration at dark was or could be
modified during technological processing and also, since the carrier
concentration is affected by the near infrared part of the
globar/mercury lamp radiation (due to persistent photoconductivity),
we estimated the carrier densities directly from the strength of
cyclotron resonance.\cite{ChiuSS76} This analysis has been performed
at higher fields, when all specimens provide a well-defined CR
absorption of Lorentzian shape. The evaluated densities for the sample
A, B and C are $n_{A,B}\sim 0.7\times 10^{11}$~cm$^{-2}$ and
$n_{C}\sim 1.8\times 10^{11}$ cm$^{-2}$, respectively.  Optionally,
this density could be further increased by illumination by an infrared
diode.  Using the CR absorption width, we roughly estimated also the
carrier mobility in samples after processing which was found to be
somewhat in excess of $10^5\unit{cm^2/(V\cdot s)}$ for all three
specimens.

%Similar mobilities were inferred also from magneto-transport measurements on
%a sample parent to C.

\section{Interpretation}

The observed departure from a single-mode CR at energy
$\hbar\omega_c$ shows a clear effect of the lateral modulation and we
discuss two possible scenarios to interpret this finding.  The first
one relies on a quantum-mechanical (single-particle) approach and
assumes the AG miniband structure due to the lateral periodic
potential. These minibands are transformed into dispersively broadened
LLs at non-zero magnetic field. The second scenario, a classical one,
recalls characteristic multimode CR absorption observed in systems
dominated by (confined) magneto-plasmons. We commence our discussion
by showing that this latter scenario is, if considered in detail,
inconsistent with our measurements.

\begin{figure}  %[htbp!]
	\begin{center}   %
   \includegraphics[width=0.8\hsize,
	angle=0]{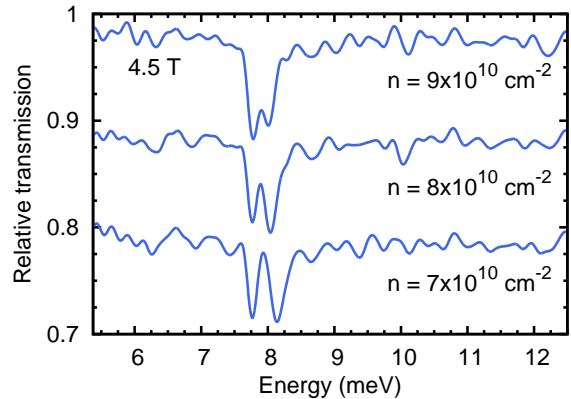}
	\caption{\small{(color online) Magneto-transmission spectra
            (shifted vertically for clarity)
            taken on
    the sample B at $B=4.5$~T with three different carrier
    concentrations adjusted by gradually increased illumination time.
    Splitting of the observed modes clearly decreases with increasing
    carrier density.}}  \label{img:6}
%The density has been subsequently increased by
%    exposing the sample to visible light for the indicated times,
%    where corresponding concentrations are 0.7$\times10^{11}$,
%    0.8$\times10^{11}$, 0.9$\times10^{11}\unit{cm^{-2}}$ for
%    additional illumination time 0, 2, 12 seconds, respectively.
%    The
%    highest illumination does not produce any increase of density
%    since the system is already saturated.
	\end{center}
\end{figure}

Splitting of the cyclotron resonance
absorption in diminishing magnetic fields, as shown in Fig.~\ref{img:3},
is reminiscent of magneto-plasma oscillations in a 2DEG.\cite{SternPRL67}
Indeed, the magneto-optical response of our
samples, which consists from a basic CR absorption line accompanied by
one or more modes at (only) higher energies, resembles spectra
taken on a 2DEG with 1D lateral modulation or on a confined
unmodulated 2DEG (stripe).\cite{MikhailovPRB05,MikhailovPRB06,FedorychIJMPB09}
Let us therefore compare our results to the response of magneto-plasmons
in a 2DEG subject to lateral modulation in both directions, in particular in
antidot lattices.\cite{KernPRL91,ZhaoAPL92} In such a case, the far
infrared or microwave magneto-absorption response includes, among
others, a characteristic lower branch located below the CR energy,
which is well documented experimentally and explained
theoretically.\cite{HeitmannPT93} This lower branch is interpreted as an
edge-magneto-plasmon (EMP), circulating around the antidot, and its
existence is not directly connected with the symmetry of the antidot
lattice.\cite{MikhailovPRB95,MikhailovPRB96}

No sign of any EMP mode below CR energy has been observed in our
experiments. We take this as a clear argument against interpretation
of the observed multi-mode CR response in terms of
magneto-plasmons. Another test to exclude magneto-plasmon effects was
performed using external illumination by an infrared diode. Upon
gradual increase of the carrier density $n$ in the sample B (as
evidenced by the strength of the CR absorption at high magnetic
fields), the distance of observed modes shown in Fig.~\ref{img:6}
decreased with $n$. For magnetoplasmonic excitations, an opposite
trend is expected.\cite{ZhaoAPL92} We also recall that the experiment
of Ref.~\onlinecite{KernPRL91} concerned deeply etched samples and
that magnetoplasmon behavior has been typically detected at
significantly higher densities.

We now turn to the discussion of experimental facts with
respect to the four criteria to observe massless Dirac fermions
in a laterally modulated 2DEG. This provides us with basic estimates on
how close or far we are from Dirac-like conditions. These criteria,
formulated in Sec.~\ref{theory}, are related to the (mini)band electronic
structure, carrier density, disorder in the system and also to the chosen
experimental technique and conditions. We first discuss criterion (i)
for which we need an estimate of the modulation potential amplitude $V_0$.
We show that infrared magneto-spectroscopy is a method suitable for this
purpose.

 \begin{figure}[t]
	\begin{center}
	\centering \includegraphics[width=0.8\hsize,
	angle=0]{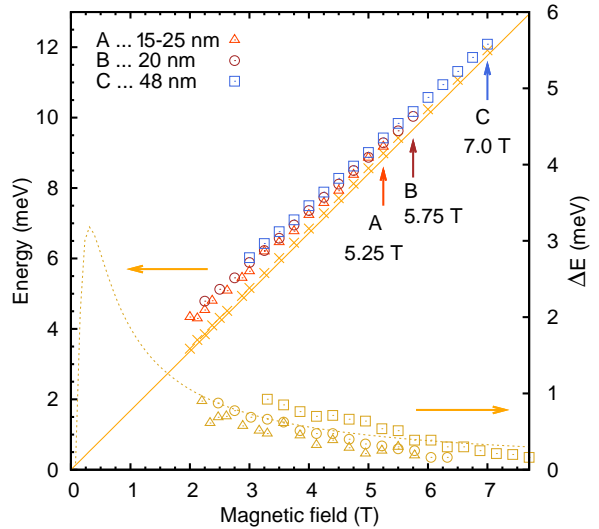} \caption{\small{(color online)
 Transitions observed in specimens A, B and C,
 cf. Fig.~\ref{img:3} (left vertical axis; depths of etched holes are
 indicated). Vertical arrows indicate
 magnetic fields at which the multi-mode character of CR absorption
 vanishes.  The position of the main CR peak in spectra taken on the
 sample A is marked by crosses.  The straight line corresponds to the
 theoretical CR-line position with an effective mass of
 $m^*=0.067m_0$, which has been derived from measurements on reference
 (unpatterned) sample (not shown).  Lower part of the
 figure (right vertical axis) shows determined values of the CR-line
 splitting $\Delta E$. The dashed line corresponds to the fit of
 $\Delta E$ for the sample A based on the theoretical model discussed
 in the text.}}
	\label{img:4}
	\end{center}
\end{figure}

The main absorption features in
Fig.~\ref{img:3} occur close to $\hbar\omega_c$ but in contrast
to the cyclotron resonance of an unmodulated 2DEG, they have an
internal structure which disappears roughly as $1/B$ (see
Fig.~\ref{img:4}) in the limit of high magnetic fields. Such behavior
suggests that the influence of the modulation potential becomes
gradually weaker with increasing $B$ and potential energy $V(\vec{r})$
acts as a perturbation to the LLs. Their spacing,
the cyclotron energy $\hbar\omega_c$, then provides the dominant
energy scale compared to the modulation potential $V_0$ and the
small parameter is $V_0/\hbar\omega_c\propto 1/B$. In the first-order
perturbation calculation,\cite{WangPRB04} the unperturbed energies
$E_n=\hbar \omega_c(n+1/2)$ become broadened into bands:
\begin{equation}\begin{array}{l}\label{eq-11}
  E_{n,\kappa_x,\kappa_y} = E_n +V_0 e^{-2\beta^2/3}L_n(4\beta^2/3)\times\\
  \hskip1cm \times\left\{2\cos
  \beta^2(\kappa_x+{\textstyle\frac{1}{\sqrt{3}}}) \cos
  \frac{\beta^2\kappa_y}{\sqrt{3}} + \cos \frac{2\beta^2
  \kappa_y}{\sqrt{3}}\right\}
  \end{array}
\end{equation}
where $\beta^2=2\pi^2\ell_0^2/a^2$, $\ell_0^2=\hbar/eB$ and
$\vec{\kappa}$ belongs to the hexagonal first magnetic Brillouin
zone. Owing to special properties of Laguerre polynomials $L_n$
\cite{Gradshteyn}, optical transition energies, that are
$E_{n+1,\kappa_x,\kappa_y}-E_{n,\kappa_x,\kappa_y}$, can be rewritten in
a simple way. Since we deal with low carrier concentrations at which
only the lowest LL is occupied, we can restrain ourselves to $n=0$,
\begin{equation}\label{eq-12}
  \Delta E_{1,0}=
  E_{1,\kappa_x,\kappa_y}-E_{0,\kappa_x,\kappa_y}=\hbar\omega_c -
  \frac{4}{3}V_0\beta^2 e^{-2\beta^2/3} b(\kappa_x,\kappa_y)
\end{equation}
where $b(\kappa_x,\kappa_y)$ denotes the curled bracket of
Eq.~(\ref{eq-11}).

\begin{figure}  %[htbp!]
  \begin{center}   %
  \includegraphics[width=0.8\hsize, angle=0]{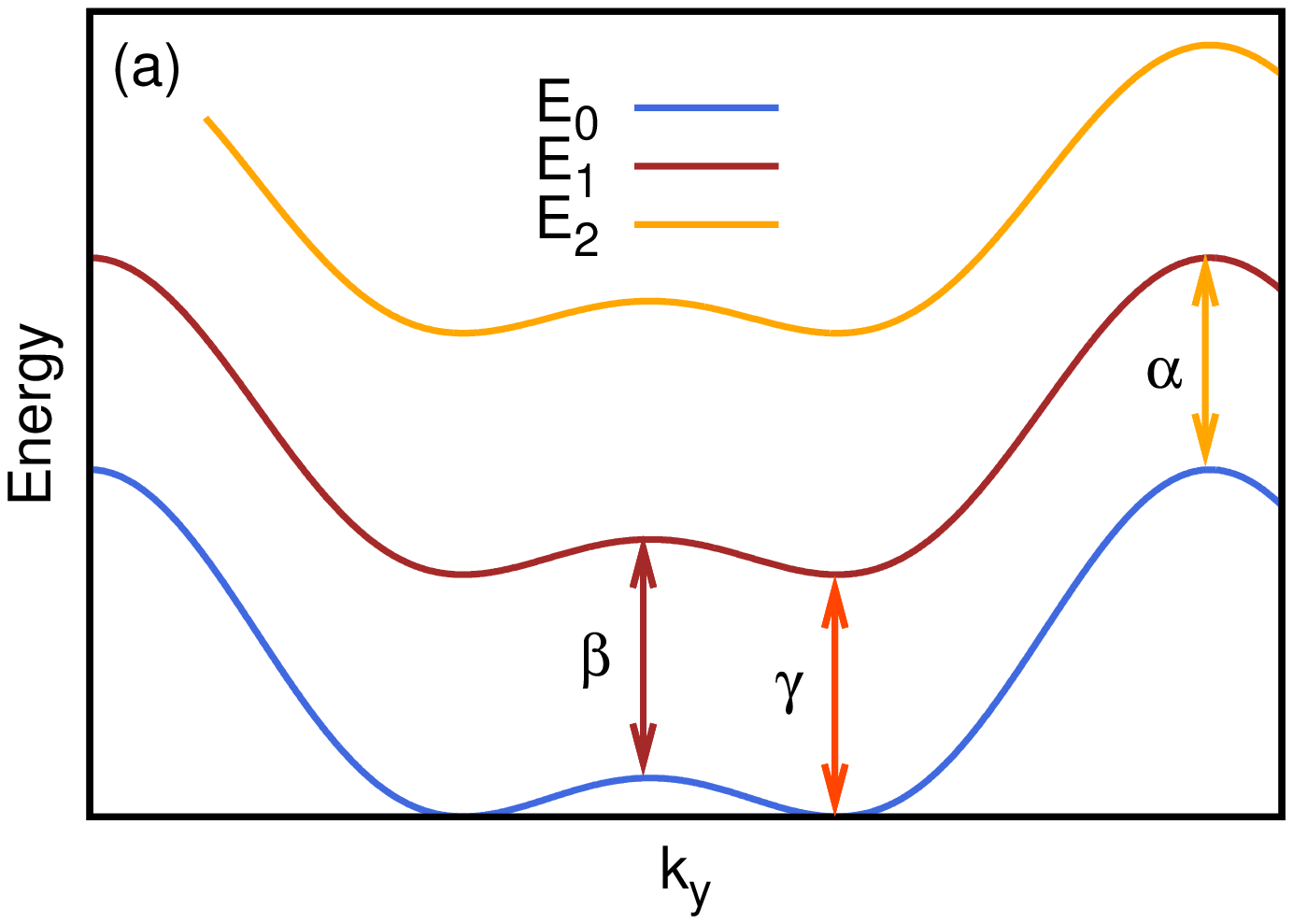}\\
	\includegraphics[width=0.8\hsize, angle=0]{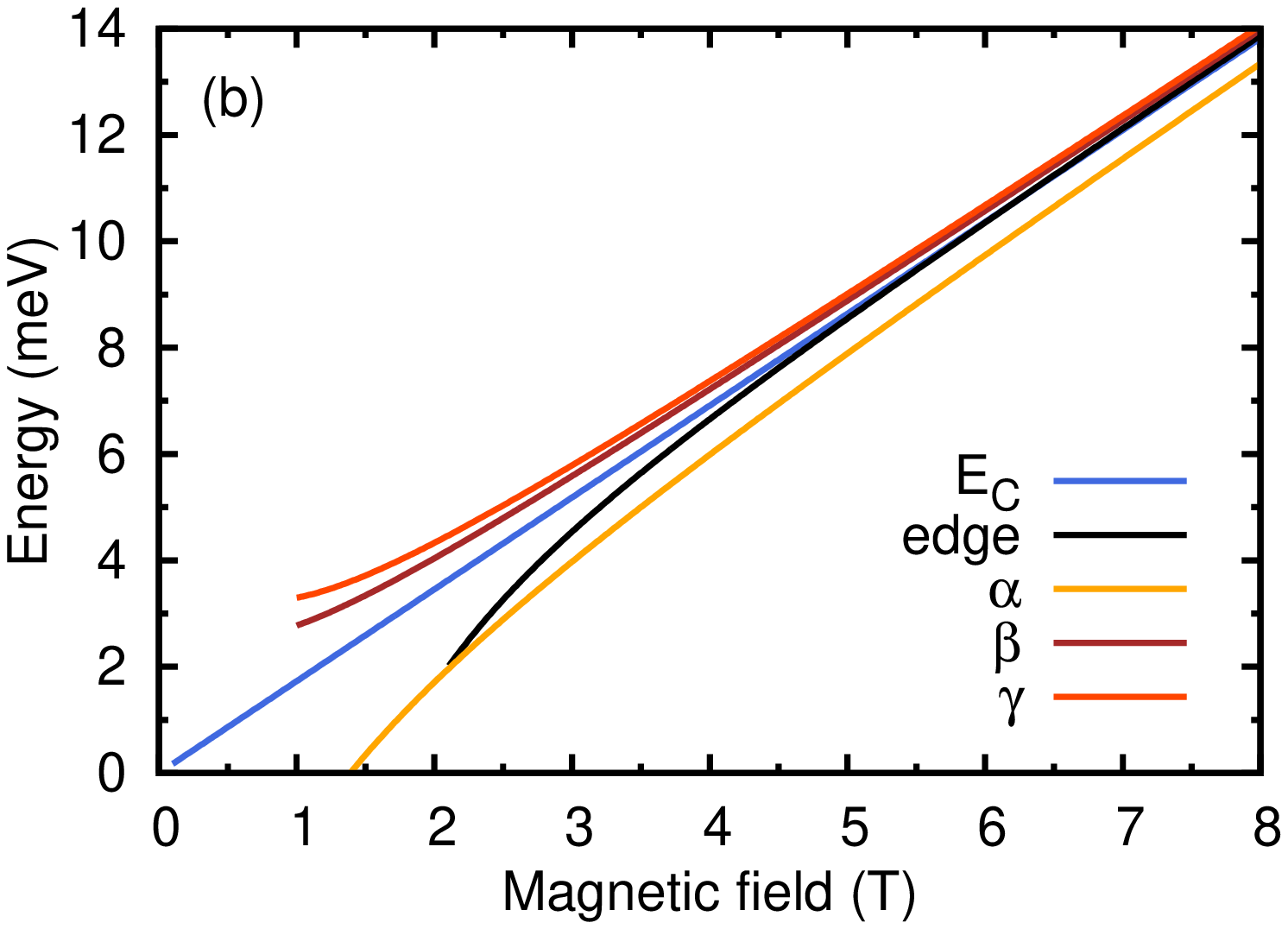}
  \end{center}
\caption{(Color online) (a) Schematic plot of three lowest lying
  LLs broadened into bands due to the lateral hexagonal
  modulation. Transitions related to the van Hove singularities in the
  joint density of states $a(\omega)$ are marked by $\alpha, \beta$
  and $\gamma$.  Whereas $\alpha$ and $\gamma$ are the band edges,
  $\beta$ is the logarithmic singularity corresponding to the saddle
  point between two neighbouring minima of $\Delta E_{10}$ in
  $\vec{\kappa}$-space.  (b) Magnetic field dependence of jDOS
  $\alpha$, $\beta$, $\gamma$ singularity positions. ``Edge'' shows
  the lowest in energy allowed transition, as defined by the position
  of the Fermi level (for $n=1.0\times 10^{11}$~cm$^{-2}$).
  $E_c=\hbar\omega_c$ is the cyclotron energy.}  \label{img:5}
\end{figure}

The optical transition energy $\Delta E_{1,0}$ enters the absorption
probability $\alpha_{1,0}$ for a photon of frequency $\omega$
that has caused transition between $n=0$ and $n=1$ LLs. Provided that
the former (latter) LL is full (empty), this probability is
proportional to\cite{note3}
\begin{equation}\label{eq-13}
  \int \frac{d^2\kappa }{(2\pi)^2}
  |\langle 1,\kappa_x,\kappa_y| p_x|0,\kappa_x,\kappa_y\rangle|^2
  \delta(\Delta E_{1,0}-\hbar\omega).
\end{equation}
If we neglect the dipole transition matrix element (transition between
the mentioned two LLs is allowed by selection rules), the
characteristic spectral features correspond to the van Hove singularities
indicated in Fig.~\ref{img:5}, in the occupation-weighted
joint density of states (jDOS):
$$ a(\omega) = \int \frac{d^2\kappa }{(2\pi)^2}\delta(\Delta
  E_{1,0}-\hbar\omega) f_{0,\kappa_x,\kappa_y}(1-f_{1,\kappa_x,\kappa_y}),
  $$
in which all those transitions at a given energy $\hbar\omega=\Delta
E_{1,0}$ count where the initial ($n=0$) state is occupied and the final
($n=1$) state is empty, as expressed by the Fermi-Dirac
factors $f$. At a filling factor $\nu=nh/eB=2$, which was assumed
in expression~(\ref{eq-13}), $a(\omega)$ is a band of the width
\begin{equation}\label{eq-14}
  w(B) = 6V_0 \beta^2 e^{-2\beta^2/3}
\end{equation}
situated close to $\omega=\omega_c$. The width of the band decreases with
decreasing $\nu$ (at constant $\beta$),
as the filling of the $n=0$ LL decreases and smaller
portions of the magnetic Brillouin zone become available for
transitions. In the limit of very large $B$, $a(\omega)$ turns into a
zero-width peak at exactly $\omega=\omega_c$.

\begin{table}[t]
  \centering
  \begin{tabular}{c|c|c|c|c}
    Sample & $d_{\rm{holes}}$ & $d_{\rm{2DEG}}$ & $V_0$ &
    $\zeta$ \\ \hline A & 15-25 & 100 & 2.2 meV &
    2.4 \\ B & 20 & 100 & 3.5 meV & 3.1 \\ C &
    48 & 115 & 4 meV & 3.6 \\
  \end{tabular}

\caption{Potential amplitude $V_0$ and the corresponding
dimensionless parameter $\zeta$ for samples A, B and C
as derived by fitting our data using Eq.~(\ref{eq-14}),
see text for details. The etching depths and the 2DEG-to-surface 
distances are also listed. }  \label{tab-01}
\end{table}

Eq.~(\ref{eq-14}) provides a reasonable basis for interpretation of
experimental data presented in Figs.~\ref{img:3}(a)-(c). The
peak-to-peak distance shown as the lower data sets in Fig.~\ref{img:4}
follows the magnetic field dependence of $w(B)$ allowing to extract
the values of $V_0$ for the particular sample. It should be noted
however, that the peak splitting observed in experiments does not
correspond to the full width $w$ as calculated using Eq.~(\ref{eq-14})
because the lower edge of the absorption band is suppressed for $\nu
<2$. This is the case of $B>2\unit{T}$ and $n<10^{11}\unit{cm^{-2}}$ when spin
degeneracy remains unresolved (Zeeman
splitting $\varepsilon_z$ is roughly an order of magnitude smaller than the 
CR peak width at $B=2$~T; $\varepsilon_z=g_se\hbar B/2m_0
\approx0.06\unit{meV}$ for $|g_s|\approx 0.44$ as appropriate in GaAs
systems\cite{Yugova:2007_a}). Features of the jDOS appearing in the
absorption band scale as $cw(B)$,
where $0<c<1$ is a constant. These features are shown in
Fig.~\ref{img:5} and correspond to the indicated transitions of the
broadened Landau bands $E_{n,\kappa_x,\kappa_y}$. The first states
that become depopulated upon the filling factor dropping below two
(that is when the magnetic field is increased) are those close to the
top of the band. Correspondingly, the transitions $\alpha$ are the first ones to
disappear from the absorption spectra.

For the remaining two features $\beta$ and $\gamma$, our form of the
potential $V(x,y)$ would imply $c=1/9$. However, since the transition
$\beta$ gives rise to a logarithmic singularity which is likely to be
smeared out, we focus on another candidate for an
absorption feature: the Fermi edge  which is also shown in the
lower panel of Fig.~\ref{img:5} (transitions from the states
close to $E_F$ to the next Landau band). Although the Fermi edge does not
precisely scale with $w(B)$, it appears at frequencies $\omega
\approx \omega_c$ hence we take $c\approx 1/3$. The values of $V_0$ inferred
from fitting our data, assuming that the splitting of the CR mode
corresponds to $\frac{1}{3}w(B)$, are shown in Tab.~\ref{tab-01}
alongside with the corresponding $\zeta$. We can now return to
criterion~(i), and see that, as
anticipated at the beginning of Sec.~\ref{exp}, shallow etching
may create modulation potential favorable for Dirac-fermion physics
in AG as quantified by the first criterion in Sec.~\ref{theory}.

While criterion (i) seems satisfied for our samples, the carrier
density is by a factor $\approx 3$ too large compared to the
requirement (ii), even in the most favourable situation (upper Dirac
cone, samples A, B). Since further lowering of $n$ may induce
metal-to-insulator transition\cite{DeSimoniAPL10} a better strategy
seems to reduce the lattice constant. Further technological
improvements will however, then be required to keep criterion (iii)
fulfilled. In our case, the measured mobility implies mean free path
only few times larger than the lattice constant and the former is
likely to deteriorate rapidly upon pushing electron lithography
closer to its limits of spatial resolution. Once the criteria (i--iii)
are met for some sample, one should proceed to low field ($B\approx
0.1\unit{T}$) and low temperature ($\lesssim 1\unit{K}$) experiments
as required by the last criterion (iv).
Since inter-LL transitions will then be in sub-THz range and
fabrication of samples homogeneous on a scale comparable to the
wavelength of absorbed light will be difficult,
magnetotransport of photoluminescence experiments seem most
promising.

\section*{Summary and Conclusions}

The concept of artificial graphene has been explored
both experimentally and theoretically. Based on a simple theoretical
model, we formulated four basic criteria that need to be met in order to
create and experimentally demonstrate the proposed graphene-like bands
in modulated semiconductor heterostructures.  We have prepared
three samples with lateral modulation and investigated them
using infrared magneto-spectroscopy. The results have been discussed
with respect to the proposed criteria with the following
conclusions. Etching an antidot (hole) lattice on the sample
surface creates lateral modulation potential with favorable strength,
which according to our model, should give
rise to a miniband structure containing
well-developed Dirac cones. An empirical rule, which connects the
depth of etched holes with the strength of the lateral potential, has
been found for our technological protocol. The
quality of prepared specimens, expressed in terms of mobility or mean
free path, could be sufficient to resolve the AG
electronic structure with the Dirac cones, nevertheless, further
increase of this quality would be desirable. The main obstacle,
preventing up to now realization of Dirac-like physics in 2DEG, seems
to be related to the interplay between the electron density and the
lattice constant. Since lowering carrier density down to
$10^9\unit{cm^{-2}}$ range seems unrealistic, we instead logically
suggest to reduce the lattice constant below 100~nm. This is
technologically challenging, given the constraints on mobility,
nevertheless still feasible.
Furthermore, we propose to focus on a higher Dirac cone, which is found
in AG minibad structure and which could be probed at higher carrier
densities, namely at $n\approx 10^{11}\unit{cm^{-2}}$ for $a=100$~nm.
With these suggestions, we believe that massless Dirac fermions can be
observed in laterally modulated 2DEG, probably using infrared or THz
magneto-spectroscopy or magneto-transport technique, in (possibly not
too distant) future.

\vspace{5mm}
\section*{Acknowledgements}

The authors would like to sincerely thank P.~Hub\'\i{}k and J.~\v
Cerm\'ak for technological assistance and J.~Wunderlich for valuable
critical remarks. Moreover, the support of the following institutions
is acknowledged: the Ministry of Education of the Czech Republic
projects LC510 and MSM0021620834, GA\v CR No. P204/10/1020, the Charles
University in Prague grants GAUK No. 425111 and SVV-2011-263306, the
Academy of Sciences of the Czech Republic via Institutional Research
Plan No. AV0Z10100521, GAAV contract KJB100100802, Fondation
\emph{NanoScience} via project Dispograph, Pr\ae mium Academi\ae{},
NSF-NEB 2020, SRC, and last but not least, EC-EuroMagNetII under
Contract No. 228043.

%\bibliography{reference}

\begin{thebibliography}{42}%
\makeatletter
\providecommand \@ifxundefined [1]{%
 \@ifx{#1\undefined}
}%
\providecommand \@ifnum [1]{%
 \ifnum #1\expandafter \@firstoftwo
 \else \expandafter \@secondoftwo
 \fi
}%
\providecommand \@ifx [1]{%
 \ifx #1\expandafter \@firstoftwo
 \else \expandafter \@secondoftwo
 \fi
}%
\providecommand \natexlab [1]{#1}%
\providecommand \enquote  [1]{``#1''}%
\providecommand \bibnamefont  [1]{#1}%
\providecommand \bibfnamefont [1]{#1}%
\providecommand \citenamefont [1]{#1}%
\providecommand \href@noop [0]{\@secondoftwo}%
\providecommand \href [0]{\begingroup \@sanitize@url \@href}%
\providecommand \@href[1]{\@@startlink{#1}\@@href}%
\providecommand \@@href[1]{\endgroup#1\@@endlink}%
\providecommand \@sanitize@url [0]{\catcode `\\12\catcode `\$12\catcode
  `\&12\catcode `\#12\catcode `\^12\catcode `\_12\catcode `\%12\relax}%
\providecommand \@@startlink[1]{}%
\providecommand \@@endlink[0]{}%
\providecommand \url  [0]{\begingroup\@sanitize@url \@url }%
\providecommand \@url [1]{\endgroup\@href {#1}{\urlprefix }}%
\providecommand \urlprefix  [0]{URL }%
\providecommand \Eprint [0]{\href }%
\providecommand \doibase [0]{http://dx.doi.org/}%
\providecommand \selectlanguage [0]{\@gobble}%
\providecommand \bibinfo  [0]{\@secondoftwo}%
\providecommand \bibfield  [0]{\@secondoftwo}%
\providecommand \translation [1]{[#1]}%
\providecommand \BibitemOpen [0]{}%
\providecommand \bibitemStop [0]{}%
\providecommand \bibitemNoStop [0]{.\EOS\space}%
\providecommand \EOS [0]{\spacefactor3000\relax}%
\providecommand \BibitemShut  [1]{\csname bibitem#1\endcsname}%
\let\auto@bib@innerbib\@empty
%</preamble>
\bibitem [{\citenamefont {Novoselov}\ \emph {et~al.}(2005)\citenamefont
  {Novoselov}, \citenamefont {Geim}, \citenamefont {Morozov}, \citenamefont
  {Jiang}, \citenamefont {Katsnelson}, \citenamefont {Grigorieva},
  \citenamefont {Dubonos},\ and\ \citenamefont {Firsov}}]{NovoselovNature05}%
  \BibitemOpen
  \bibfield  {author} {\bibinfo {author} {\bibfnamefont {K.~S.}\ \bibnamefont
  {Novoselov}}, \bibinfo {author} {\bibfnamefont {A.~K.}\ \bibnamefont {Geim}},
  \bibinfo {author} {\bibfnamefont {S.~V.}\ \bibnamefont {Morozov}}, \bibinfo
  {author} {\bibfnamefont {D.}~\bibnamefont {Jiang}}, \bibinfo {author}
  {\bibfnamefont {M.~I.}\ \bibnamefont {Katsnelson}}, \bibinfo {author}
  {\bibfnamefont {I.~V.}\ \bibnamefont {Grigorieva}}, \bibinfo {author}
  {\bibfnamefont {S.~V.}\ \bibnamefont {Dubonos}}, \ and\ \bibinfo {author}
  {\bibfnamefont {A.~A.}\ \bibnamefont {Firsov}},\ }\href@noop {} {\bibfield
  {journal} {\bibinfo  {journal} {Nature}\ }\textbf {\bibinfo {volume} {438}},\
  \bibinfo {pages} {197} (\bibinfo {year} {2005})}\BibitemShut {NoStop}%
\bibitem [{\citenamefont {Zhang}\ \emph {et~al.}(2005)\citenamefont {Zhang},
  \citenamefont {Tan}, \citenamefont {Stormer},\ and\ \citenamefont
  {Kim}}]{ZhangNature05}%
  \BibitemOpen
  \bibfield  {author} {\bibinfo {author} {\bibfnamefont {Y.~B.}\ \bibnamefont
  {Zhang}}, \bibinfo {author} {\bibfnamefont {Y.~W.}\ \bibnamefont {Tan}},
  \bibinfo {author} {\bibfnamefont {H.~L.}\ \bibnamefont {Stormer}}, \ and\
  \bibinfo {author} {\bibfnamefont {P.}~\bibnamefont {Kim}},\ }\href@noop {}
  {\bibfield  {journal} {\bibinfo  {journal} {Nature}\ }\textbf {\bibinfo
  {volume} {438}},\ \bibinfo {pages} {201} (\bibinfo {year}
  {2005})}\BibitemShut {NoStop}%
\bibitem [{\citenamefont {Geim}\ and\ \citenamefont
  {Novoselov}(2007)}]{GeimNatureMaterial07}%
  \BibitemOpen
  \bibfield  {author} {\bibinfo {author} {\bibfnamefont {A.~K.}\ \bibnamefont
  {Geim}}\ and\ \bibinfo {author} {\bibfnamefont {K.~S.}\ \bibnamefont
  {Novoselov}},\ }\href@noop {} {\bibfield  {journal} {\bibinfo  {journal}
  {Nature Mater.}\ }\textbf {\bibinfo {volume} {6}},\ \bibinfo {pages} {183}
  (\bibinfo {year} {2007})}\BibitemShut {NoStop}%
\bibitem [{\citenamefont {Grynberg}\ \emph {et~al.}(1993)\citenamefont
  {Grynberg}, \citenamefont {Lounis}, \citenamefont {Verkerk}, \citenamefont
  {Courtois},\ and\ \citenamefont {Salomon}}]{GrynbergPRL93}%
  \BibitemOpen
  \bibfield  {author} {\bibinfo {author} {\bibfnamefont {G.}~\bibnamefont
  {Grynberg}}, \bibinfo {author} {\bibfnamefont {B.}~\bibnamefont {Lounis}},
  \bibinfo {author} {\bibfnamefont {P.}~\bibnamefont {Verkerk}}, \bibinfo
  {author} {\bibfnamefont {J.-Y.}\ \bibnamefont {Courtois}}, \ and\ \bibinfo
  {author} {\bibfnamefont {C.}~\bibnamefont {Salomon}},\ }\href@noop {}
  {\bibfield  {journal} {\bibinfo  {journal} {Phys. Rev. Lett.}\ }\textbf
  {\bibinfo {volume} {70}},\ \bibinfo {pages} {2249} (\bibinfo {year}
  {1993})}\BibitemShut {NoStop}%
\bibitem [{\citenamefont {Zhu}\ \emph {et~al.}(2007)\citenamefont {Zhu},
  \citenamefont {Wang},\ and\ \citenamefont {Duan}}]{ZhuPRL07}%
  \BibitemOpen
  \bibfield  {author} {\bibinfo {author} {\bibfnamefont {S.-L.}\ \bibnamefont
  {Zhu}}, \bibinfo {author} {\bibfnamefont {B.}~\bibnamefont {Wang}}, \ and\
  \bibinfo {author} {\bibfnamefont {L.-M.}\ \bibnamefont {Duan}},\ }\href
  {\doibase 10.1103/PhysRevLett.98.260402} {\bibfield  {journal} {\bibinfo
  {journal} {Phys. Rev. Lett.}\ }\textbf {\bibinfo {volume} {98}},\ \bibinfo
  {pages} {260402} (\bibinfo {year} {2007})}\BibitemShut {NoStop}%
\bibitem [{\citenamefont {Wunsch}\ \emph {et~al.}(2008)\citenamefont {Wunsch},
  \citenamefont {Guinea},\ and\ \citenamefont {Sols}}]{WunchNJP08}%
  \BibitemOpen
  \bibfield  {author} {\bibinfo {author} {\bibfnamefont {B.}~\bibnamefont
  {Wunsch}}, \bibinfo {author} {\bibfnamefont {F.}~\bibnamefont {Guinea}}, \
  and\ \bibinfo {author} {\bibfnamefont {F.}~\bibnamefont {Sols}},\ }\href@noop
  {} {\bibfield  {journal} {\bibinfo  {journal} {New Journal of Physics}\
  }\textbf {\bibinfo {volume} {10}},\ \bibinfo {pages} {103027} (\bibinfo
  {year} {2008})}\BibitemShut {NoStop}%
\bibitem [{\citenamefont {Park}\ \emph {et~al.}(2008)\citenamefont {Park},
  \citenamefont {Yang}, \citenamefont {Son}, \citenamefont {Cohen},\ and\
  \citenamefont {Louie}}]{ParkPRL08}%
  \BibitemOpen
  \bibfield  {author} {\bibinfo {author} {\bibfnamefont {C.-H.}\ \bibnamefont
  {Park}}, \bibinfo {author} {\bibfnamefont {L.}~\bibnamefont {Yang}}, \bibinfo
  {author} {\bibfnamefont {Y.-W.}\ \bibnamefont {Son}}, \bibinfo {author}
  {\bibfnamefont {M.~L.}\ \bibnamefont {Cohen}}, \ and\ \bibinfo {author}
  {\bibfnamefont {S.~G.}\ \bibnamefont {Louie}},\ }\href {\doibase
  10.1103/PhysRevLett.101.126804} {\bibfield  {journal} {\bibinfo  {journal}
  {Phys. Rev. Lett.}\ }\textbf {\bibinfo {volume} {101}},\ \bibinfo {pages}
  {126804} (\bibinfo {year} {2008})}\BibitemShut {NoStop}%
\bibitem [{\citenamefont {Park}\ and\ \citenamefont {Louie}(2009)}]{ParkNL09}%
  \BibitemOpen
  \bibfield  {author} {\bibinfo {author} {\bibfnamefont {C.-H.}\ \bibnamefont
  {Park}}\ and\ \bibinfo {author} {\bibfnamefont {S.~G.}\ \bibnamefont
  {Louie}},\ }\href@noop {} {\bibfield  {journal} {\bibinfo  {journal} {Nano
  Letters}\ }\textbf {\bibinfo {volume} {9}},\ \bibinfo {pages} {1793}
  (\bibinfo {year} {2009})}\BibitemShut {NoStop}%
\bibitem [{\citenamefont {Gibertini}\ \emph {et~al.}(2009)\citenamefont
  {Gibertini}, \citenamefont {Singha}, \citenamefont {Pellegrini},
  \citenamefont {Polini}, \citenamefont {Vignale}, \citenamefont {Pinczuk},
  \citenamefont {Pfeiffer},\ and\ \citenamefont {West}}]{GibertiniPRB09}%
  \BibitemOpen
  \bibfield  {author} {\bibinfo {author} {\bibfnamefont {M.}~\bibnamefont
  {Gibertini}}, \bibinfo {author} {\bibfnamefont {A.}~\bibnamefont {Singha}},
  \bibinfo {author} {\bibfnamefont {V.}~\bibnamefont {Pellegrini}}, \bibinfo
  {author} {\bibfnamefont {M.}~\bibnamefont {Polini}}, \bibinfo {author}
  {\bibfnamefont {G.}~\bibnamefont {Vignale}}, \bibinfo {author} {\bibfnamefont
  {A.}~\bibnamefont {Pinczuk}}, \bibinfo {author} {\bibfnamefont {L.~N.}\
  \bibnamefont {Pfeiffer}}, \ and\ \bibinfo {author} {\bibfnamefont {K.~W.}\
  \bibnamefont {West}},\ }\href {\doibase 10.1103/PhysRevB.79.241406}
  {\bibfield  {journal} {\bibinfo  {journal} {Phys. Rev. B}\ }\textbf {\bibinfo
  {volume} {79}},\ \bibinfo {pages} {241406} (\bibinfo {year}
  {2009})}\BibitemShut {NoStop}%
\bibitem [{\citenamefont {{Dubois, S. M.-M.}}\ \emph
  {et~al.}(2009)\citenamefont {{Dubois, S. M.-M.}}, \citenamefont {{Zanolli,
  Z.}}, \citenamefont {{Declerck, X.}},\ and\ \citenamefont {{Charlier,
  J.-C.}}}]{DuboisEPJB09}%
  \BibitemOpen
  \bibfield  {author} {\bibinfo {author} {\bibnamefont {{Dubois, S. M.-M.}}},
  \bibinfo {author} {\bibnamefont {{Zanolli, Z.}}}, \bibinfo {author}
  {\bibnamefont {{Declerck, X.}}}, \ and\ \bibinfo {author} {\bibnamefont
  {{Charlier, J.-C.}}},\ }\href {\doibase 10.1140/epjb/e2009-00327-8}
  {\bibfield  {journal} {\bibinfo  {journal} {Eur. Phys. J. B}\ }\textbf
  {\bibinfo {volume} {72}},\ \bibinfo {pages} {1} (\bibinfo {year}
  {2009})}\BibitemShut {NoStop}%
\bibitem [{\citenamefont {Rycerz}\ \emph {et~al.}(2007)\citenamefont {Rycerz},
  \citenamefont {{Tworzyd{\l}o}},\ and\ \citenamefont
  {Beenakker}}]{RycerzNaturePhys07}%
  \BibitemOpen
  \bibfield  {author} {\bibinfo {author} {\bibfnamefont {A.}~\bibnamefont
  {Rycerz}}, \bibinfo {author} {\bibfnamefont {J.}~\bibnamefont
  {{Tworzyd{\l}o}}}, \ and\ \bibinfo {author} {\bibfnamefont {C.~W.~J.}\
  \bibnamefont {Beenakker}},\ }\href@noop {} {\bibfield  {journal} {\bibinfo
  {journal} {Nature Phys.}\ }\textbf {\bibinfo {volume} {3}},\ \bibinfo {pages}
  {172 } (\bibinfo {year} {2007})}\BibitemShut {NoStop}%
\bibitem [{\citenamefont {Cheianov}\ \emph {et~al.}(2007)\citenamefont
  {Cheianov}, \citenamefont {Fal'ko},\ and\ \citenamefont
  {Altshuler}}]{CheianovScience07}%
  \BibitemOpen
  \bibfield  {author} {\bibinfo {author} {\bibfnamefont {V.~V.}\ \bibnamefont
  {Cheianov}}, \bibinfo {author} {\bibfnamefont {V.}~\bibnamefont {Fal'ko}}, \
  and\ \bibinfo {author} {\bibfnamefont {B.~L.}\ \bibnamefont {Altshuler}},\
  }\href {\doibase 10.1126/science.1138020} {\bibfield  {journal} {\bibinfo
  {journal} {Science}\ }\textbf {\bibinfo {volume} {315}},\ \bibinfo {pages}
  {1252} (\bibinfo {year} {2007})}\BibitemShut {NoStop}%
\bibitem [{\citenamefont {Garcia-Pomar}\ \emph {et~al.}(2008)\citenamefont
  {Garcia-Pomar}, \citenamefont {Cortijo},\ and\ \citenamefont
  {Nieto-Vesperinas}}]{Garcia-PomarPRL08}%
  \BibitemOpen
  \bibfield  {author} {\bibinfo {author} {\bibfnamefont {J.~L.}\ \bibnamefont
  {Garcia-Pomar}}, \bibinfo {author} {\bibfnamefont {A.}~\bibnamefont
  {Cortijo}}, \ and\ \bibinfo {author} {\bibfnamefont {M.}~\bibnamefont
  {Nieto-Vesperinas}},\ }\href {\doibase 10.1103/PhysRevLett.100.236801}
  {\bibfield  {journal} {\bibinfo  {journal} {Phys. Rev. Lett.}\ }\textbf
  {\bibinfo {volume} {100}},\ \bibinfo {pages} {236801} (\bibinfo {year}
  {2008})}\BibitemShut {NoStop}%

\bibitem{Eigler:1990_a} D.M.~Eigler and E.K.~Schweizer, % doi: 10.1038/344524a0
  Nature {\bf 344}, 524 (1990).

\bibitem [{\citenamefont {Kern}\ \emph {et~al.}(1991)\citenamefont {Kern},
  \citenamefont {Heitmann}, \citenamefont {Grambow}, \citenamefont {Zhang},\
  and\ \citenamefont {Ploog}}]{KernPRL91}%
  \BibitemOpen
  \bibfield  {author} {\bibinfo {author} {\bibfnamefont {K.}~\bibnamefont
  {Kern}}, \bibinfo {author} {\bibfnamefont {D.}~\bibnamefont {Heitmann}},
  \bibinfo {author} {\bibfnamefont {P.}~\bibnamefont {Grambow}}, \bibinfo
  {author} {\bibfnamefont {Y.~H.}\ \bibnamefont {Zhang}}, \ and\ \bibinfo
  {author} {\bibfnamefont {K.}~\bibnamefont {Ploog}},\ }\href@noop {}
  {\bibfield  {journal} {\bibinfo  {journal} {Phys. Rev. Lett.}\ }\textbf
  {\bibinfo {volume} {66}},\ \bibinfo {pages} {1618} (\bibinfo {year}
  {1991})}\BibitemShut {NoStop}%
\bibitem [{\citenamefont {Weiss}\ \emph {et~al.}(1989)\citenamefont {Weiss},
  \citenamefont {Klitzing}, \citenamefont {Ploog},\ and\ \citenamefont
  {Weimann}}]{WeissEPL89}%
  \BibitemOpen
  \bibfield  {author} {\bibinfo {author} {\bibfnamefont {D.}~\bibnamefont
  {Weiss}}, \bibinfo {author} {\bibfnamefont {K.~V.}\ \bibnamefont {Klitzing}},
  \bibinfo {author} {\bibfnamefont {K.}~\bibnamefont {Ploog}}, \ and\ \bibinfo
  {author} {\bibfnamefont {G.}~\bibnamefont {Weimann}},\ }\href
  {http://stacks.iop.org/0295-5075/8/i=2/a=012} {\bibfield  {journal} {\bibinfo
   {journal} {EPL (Europhysics Letters)}\ }\textbf {\bibinfo {volume} {8}},\
  \bibinfo {pages} {179} (\bibinfo {year} {1989})}\BibitemShut {NoStop}%
\bibitem [{\citenamefont {Mikhailov}(1996)}]{MikhailovPRB96}%
  \BibitemOpen
  \bibfield  {author} {\bibinfo {author} {\bibfnamefont {S.~A.}\ \bibnamefont
  {Mikhailov}},\ }\href@noop {} {\bibfield  {journal} {\bibinfo  {journal}
  {Phys. Rev. B}\ }\textbf {\bibinfo {volume} {54}},\ \bibinfo {pages} {R14293}
  (\bibinfo {year} {1996})}\BibitemShut {NoStop}%
\bibitem [{\citenamefont {Beenakker}(1989)}]{BeenakkerPRL89}%
  \BibitemOpen
  \bibfield  {author} {\bibinfo {author} {\bibfnamefont {C.~W.~J.}\
  \bibnamefont {Beenakker}},\ }\href {\doibase 10.1103/PhysRevLett.62.2020}
  {\bibfield  {journal} {\bibinfo  {journal} {Phys. Rev. Lett.}\ }\textbf
  {\bibinfo {volume} {62}},\ \bibinfo {pages} {2020} (\bibinfo {year}
  {1989})}\BibitemShut {NoStop}%
\bibitem [{\citenamefont {{St\v{r}eda}}\ and\ \citenamefont
  {MacDonald}(1990)}]{StredaPRB90}%
  \BibitemOpen
  \bibfield  {author} {\bibinfo {author} {\bibfnamefont {P.}~\bibnamefont
  {{St\v{r}eda}}}\ and\ \bibinfo {author} {\bibfnamefont {A.~H.}\ \bibnamefont
  {MacDonald}},\ }\href {\doibase 10.1103/PhysRevB.41.11892} {\bibfield
  {journal} {\bibinfo  {journal} {Phys. Rev. B}\ }\textbf {\bibinfo {volume}
  {41}},\ \bibinfo {pages} {11892} (\bibinfo {year} {1990})}
  \BibitemShut {NoStop}%

\bibitem{Gvozdikov:2007_a} V.M. Gvozdikov,
  Phys. Rev. B {\bf 75}, 115106 (2007).

\bibitem [{\citenamefont {Albrecht}\ \emph {et~al.}(1999)\citenamefont
  {Albrecht}, \citenamefont {Smet}, \citenamefont {Weiss}, \citenamefont {von
  Klitzing}, \citenamefont {Hennig}, \citenamefont {Langenbuch}, \citenamefont
  {Suhrke}, \citenamefont {R\"ossler}, \citenamefont {Umansky},\ and\
  \citenamefont {Schweizer}}]{AlbrechtPRL99}%
  \BibitemOpen
  \bibfield  {author} {\bibinfo {author} {\bibfnamefont {C.}~\bibnamefont
  {Albrecht}}, \bibinfo {author} {\bibfnamefont {J.~H.}\ \bibnamefont {Smet}},
  \bibinfo {author} {\bibfnamefont {D.}~\bibnamefont {Weiss}}, \bibinfo
  {author} {\bibfnamefont {K.}~\bibnamefont {von Klitzing}}, \bibinfo {author}
  {\bibfnamefont {R.}~\bibnamefont {Hennig}}, \bibinfo {author} {\bibfnamefont
  {M.}~\bibnamefont {Langenbuch}}, \bibinfo {author} {\bibfnamefont
  {M.}~\bibnamefont {Suhrke}}, \bibinfo {author} {\bibfnamefont
  {U.}~\bibnamefont {R\"ossler}}, \bibinfo {author} {\bibfnamefont
  {V.}~\bibnamefont {Umansky}}, \ and\ \bibinfo {author} {\bibfnamefont
  {H.}~\bibnamefont {Schweizer}},\ }\href {\doibase
  10.1103/PhysRevLett.83.2234} {\bibfield  {journal} {\bibinfo  {journal}
  {Phys. Rev. Lett.}\ }\textbf {\bibinfo {volume} {83}},\ \bibinfo {pages}
  {2234} (\bibinfo {year} {1999})}\BibitemShut {NoStop}%
\bibitem [{\citenamefont {Olszewski}\ \emph {et~al.}(2004)\citenamefont
  {Olszewski}, \citenamefont {Pietrachowicz},\ and\ \citenamefont
  {Baszczak}}]{OlszewskiPSSB04}%
  \BibitemOpen
  \bibfield  {author} {\bibinfo {author} {\bibfnamefont {S.}~\bibnamefont
  {Olszewski}}, \bibinfo {author} {\bibfnamefont {M.}~\bibnamefont
  {Pietrachowicz}}, \ and\ \bibinfo {author} {\bibfnamefont {M.}~\bibnamefont
  {Baszczak}},\ }\href@noop {} {\bibfield  {journal} {\bibinfo  {journal}
  {phys. stat. sol. (b)}\ }\textbf {\bibinfo {volume} {241}},\ \bibinfo {pages}
  {3572} (\bibinfo {year} {2004})}\BibitemShut {NoStop}%

\bibitem{Gvozdikov:2007_b} V.M.~Gvozdikov,
  Phys. Rev. B {\bf 76}, 235125 (2007).

\bibitem [{\citenamefont {Hofstadter}(1976)}]{HofstadterPRB76}%
  \BibitemOpen
  \bibfield  {author} {\bibinfo {author} {\bibfnamefont {D.~R.}\ \bibnamefont
  {Hofstadter}},\ }\href@noop {} {\bibfield  {journal} {\bibinfo  {journal}
  {Phys. Rev. B}\ }\textbf {\bibinfo {volume} {14}},\ \bibinfo {pages} {2239}
  (\bibinfo {year} {1976})}\BibitemShut {NoStop}%
\bibitem [{\citenamefont {Geisler}\ \emph {et~al.}(2004)\citenamefont
  {Geisler}, \citenamefont {Smet}, \citenamefont {Umansky}, \citenamefont {von
  Klitzing}, \citenamefont {Naundorf}, \citenamefont {Ketzmerick},\ and\
  \citenamefont {Schweizer}}]{GeislerPRL04}%
  \BibitemOpen
  \bibfield  {author} {\bibinfo {author} {\bibfnamefont {M.~C.}\ \bibnamefont
  {Geisler}}, \bibinfo {author} {\bibfnamefont {J.~H.}\ \bibnamefont {Smet}},
  \bibinfo {author} {\bibfnamefont {V.}~\bibnamefont {Umansky}}, \bibinfo
  {author} {\bibfnamefont {K.}~\bibnamefont {von Klitzing}}, \bibinfo {author}
  {\bibfnamefont {B.}~\bibnamefont {Naundorf}}, \bibinfo {author}
  {\bibfnamefont {R.}~\bibnamefont {Ketzmerick}}, \ and\ \bibinfo {author}
  {\bibfnamefont {H.}~\bibnamefont {Schweizer}},\ }\href {\doibase
  10.1103/PhysRevLett.92.256801} {\bibfield  {journal} {\bibinfo  {journal}
  {Phys. Rev. Lett.}\ }\textbf {\bibinfo {volume} {92}},\ \bibinfo {pages}
  {256801} (\bibinfo {year} {2004})}\BibitemShut {NoStop}%

\bibitem{Pfannkuche:1992_a} D.~Pfannkuche and R.R.~Gerhardts,
  Phys. Rev. B {\bf 46}, 12606 (1992).

\bibitem [{\citenamefont {{De Simoni}}\ \emph {et~al.}(2010)\citenamefont {{De
  Simoni}}, \citenamefont {Singha}, \citenamefont {Gibertini}, \citenamefont
  {Karmakar}, \citenamefont {Polini}, \citenamefont {Piazza}, \citenamefont
  {Pfeiffer}, \citenamefont {West}, \citenamefont {Beltram},\ and\
  \citenamefont {Pellegrini}}]{DeSimoniAPL10}%
  \BibitemOpen
  \bibfield  {author} {\bibinfo {author} {\bibfnamefont {G.}~\bibnamefont {{De
  Simoni}}}, \bibinfo {author} {\bibfnamefont {A.}~\bibnamefont {Singha}},
  \bibinfo {author} {\bibfnamefont {M.}~\bibnamefont {Gibertini}}, \bibinfo
  {author} {\bibfnamefont {B.}~\bibnamefont {Karmakar}}, \bibinfo {author}
  {\bibfnamefont {M.}~\bibnamefont {Polini}}, \bibinfo {author} {\bibfnamefont
  {V.}~\bibnamefont {Piazza}}, \bibinfo {author} {\bibfnamefont {L.~N.}\
  \bibnamefont {Pfeiffer}}, \bibinfo {author} {\bibfnamefont {K.~W.}\
  \bibnamefont {West}}, \bibinfo {author} {\bibfnamefont {F.}~\bibnamefont
  {Beltram}}, \ and\ \bibinfo {author} {\bibfnamefont {V.}~\bibnamefont
  {Pellegrini}},\ }\href {\doibase DOI:10.1063/1.3493189} {\bibfield  {journal}
  {\bibinfo  {journal} {Appl. Phys. Lett.}\ }\textbf {\bibinfo {volume} {97}},\
  \bibinfo {pages} {132113} (\bibinfo {year} {2010})}\BibitemShut {NoStop}%
\bibitem [{\citenamefont {Singha}\ \emph {et~al.}(2011)\citenamefont {Singha},
  \citenamefont {Gibertini}, \citenamefont {Karmakar}, \citenamefont {Yuan},
  \citenamefont {Polini}, \citenamefont {Vignale}, \citenamefont {Katsnelson},
  \citenamefont {Pinczuk}, \citenamefont {Pfeiffer}, \citenamefont {West},\
  and\ \citenamefont {Pellegrini}}]{SinghaScience11}%
  \BibitemOpen
  \bibfield  {author} {\bibinfo {author} {\bibfnamefont {A.}~\bibnamefont
  {Singha}}, \bibinfo {author} {\bibfnamefont {M.}~\bibnamefont {Gibertini}},
  \bibinfo {author} {\bibfnamefont {B.}~\bibnamefont {Karmakar}}, \bibinfo
  {author} {\bibfnamefont {S.}~\bibnamefont {Yuan}}, \bibinfo {author}
  {\bibfnamefont {M.}~\bibnamefont {Polini}}, \bibinfo {author} {\bibfnamefont
  {G.}~\bibnamefont {Vignale}}, \bibinfo {author} {\bibfnamefont {M.~I.}\
  \bibnamefont {Katsnelson}}, \bibinfo {author} {\bibfnamefont
  {A.}~\bibnamefont {Pinczuk}}, \bibinfo {author} {\bibfnamefont {L.~N.}\
  \bibnamefont {Pfeiffer}}, \bibinfo {author} {\bibfnamefont {K.~W.}\
  \bibnamefont {West}}, \ and\ \bibinfo {author} {\bibfnamefont
  {V.}~\bibnamefont {Pellegrini}},\ }\href {\doibase 10.1126/science.1204333}
  {\bibfield  {journal} {\bibinfo  {journal} {Science}\ }\textbf {\bibinfo
  {volume} {332}},\ \bibinfo {pages} {1176} (\bibinfo {year}
  {2011})}\BibitemShut {NoStop}%
\bibitem [{\citenamefont {Sadowski}\ \emph {et~al.}(2006)\citenamefont
  {Sadowski}, \citenamefont {Martinez}, \citenamefont {Potemski}, \citenamefont
  {Berger},\ and\ \citenamefont {de~Heer}}]{SadowskiPRL06}%
  \BibitemOpen
  \bibfield  {author} {\bibinfo {author} {\bibfnamefont {M.~L.}\ \bibnamefont
  {Sadowski}}, \bibinfo {author} {\bibfnamefont {G.}~\bibnamefont {Martinez}},
  \bibinfo {author} {\bibfnamefont {M.}~\bibnamefont {Potemski}}, \bibinfo
  {author} {\bibfnamefont {C.}~\bibnamefont {Berger}}, \ and\ \bibinfo {author}
  {\bibfnamefont {W.~A.}\ \bibnamefont {de~Heer}},\ }\href@noop {} {\bibfield
  {journal} {\bibinfo  {journal} {Phys. Rev. Lett.}\ }\textbf {\bibinfo
  {volume} {97}},\ \bibinfo {pages} {266405} (\bibinfo {year}
  {2006})}\BibitemShut {NoStop}%
\bibitem [{\citenamefont {Jiang}\ \emph {et~al.}(2007)\citenamefont {Jiang},
  \citenamefont {Henriksen}, \citenamefont {Tung}, \citenamefont {Wang},
  \citenamefont {Schwartz}, \citenamefont {Han}, \citenamefont {Kim},\ and\
  \citenamefont {Stormer}}]{JiangPRL07}%
  \BibitemOpen
  \bibfield  {author} {\bibinfo {author} {\bibfnamefont {Z.}~\bibnamefont
  {Jiang}}, \bibinfo {author} {\bibfnamefont {E.~A.}\ \bibnamefont
  {Henriksen}}, \bibinfo {author} {\bibfnamefont {L.~C.}\ \bibnamefont {Tung}},
  \bibinfo {author} {\bibfnamefont {Y.-J.}\ \bibnamefont {Wang}}, \bibinfo
  {author} {\bibfnamefont {M.~E.}\ \bibnamefont {Schwartz}}, \bibinfo {author}
  {\bibfnamefont {M.~Y.}\ \bibnamefont {Han}}, \bibinfo {author} {\bibfnamefont
  {P.}~\bibnamefont {Kim}}, \ and\ \bibinfo {author} {\bibfnamefont {H.~L.}\
  \bibnamefont {Stormer}},\ }\href {\doibase 10.1103/PhysRevLett.98.197403}
  {\bibfield  {journal} {\bibinfo  {journal} {Phys. Rev. Lett.}\ }\textbf
  {\bibinfo {volume} {98}},\ \bibinfo {pages} {197403} (\bibinfo {year}
  {2007})}\BibitemShut {NoStop}%
\bibitem [{\citenamefont {Deacon}\ \emph {et~al.}(2007)\citenamefont {Deacon},
  \citenamefont {Chuang}, \citenamefont {Nicholas}, \citenamefont {Novoselov},\
  and\ \citenamefont {Geim}}]{DeaconPRB07}%
  \BibitemOpen
  \bibfield  {author} {\bibinfo {author} {\bibfnamefont {R.~S.}\ \bibnamefont
  {Deacon}}, \bibinfo {author} {\bibfnamefont {K.-C.}\ \bibnamefont {Chuang}},
  \bibinfo {author} {\bibfnamefont {R.~J.}\ \bibnamefont {Nicholas}}, \bibinfo
  {author} {\bibfnamefont {K.~S.}\ \bibnamefont {Novoselov}}, \ and\ \bibinfo
  {author} {\bibfnamefont {A.~K.}\ \bibnamefont {Geim}},\ }\href {\doibase
  10.1103/PhysRevB.76.081406} {\bibfield  {journal} {\bibinfo  {journal} {Phys.
  Rev. B}\ }\textbf {\bibinfo {volume} {76}},\ \bibinfo {pages} {081406}
  (\bibinfo {year} {2007})}\BibitemShut {NoStop}%

\bibitem{note1} 
Some previous calculations\cite{ParkNL09,GibertiniPRB09} used
muffin-tin potential whose parameters are (at minimum) three: antidot
distance, size and depth.  Realistic form of the potential in an
experimental sample is naturally hard to ascertain in detail, hence
the advantage of the present model is that it is both generic and
simple (single-parametric), allowing to classify types of miniband
structures that may arise while not compromising the model
validity. The lesson from studies with muffin-tin potential however is
that higher harmonics added to $V(\vec{r})$ do not ruin the Dirac
cones by opening gaps, provided such additional terms preserve the
honeycomb symmetry. We thank Rafa\l{} Oszwa\l dowski for his comments
on this topic.

\bibitem{note2} 
Evolution of wavefunctions (modulus square) upon moving on a small
circle in momentum space around the K-point was
studied. Degenerate-level perturbative results\cite{ParkNL09} were
recovered for both Dirac cones even in the presence of mixing to
levels far from the cones.

\bibitem [{\citenamefont {Soibel}\ \emph {et~al.}(1996)\citenamefont {Soibel},
  \citenamefont {Meirav}, \citenamefont {Mahalu},\ and\ \citenamefont
  {Shtrikman}}]{SoibelSST96}%
  \BibitemOpen
  \bibfield  {author} {\bibinfo {author} {\bibfnamefont {A.}~\bibnamefont
  {Soibel}}, \bibinfo {author} {\bibfnamefont {U.}~\bibnamefont {Meirav}},
  \bibinfo {author} {\bibfnamefont {D.}~\bibnamefont {Mahalu}}, \ and\ \bibinfo
  {author} {\bibfnamefont {H.}~\bibnamefont {Shtrikman}},\ }\href@noop {}
  {\bibfield  {journal} {\bibinfo  {journal} {Semiconductor Science and
  Technology}\ }\textbf {\bibinfo {volume} {11}},\ \bibinfo {pages} {1756}
  (\bibinfo {year} {1996})}\BibitemShut {NoStop}%
\bibitem [{\citenamefont {Hugger}\ \emph {et~al.}(2008)\citenamefont {Hugger},
  \citenamefont {Heinzel},\ and\ \citenamefont {Thurn-Albrecht}}]{HuggerAPL08}%
  \BibitemOpen
  \bibfield  {author} {\bibinfo {author} {\bibfnamefont {S.}~\bibnamefont
  {Hugger}}, \bibinfo {author} {\bibfnamefont {T.}~\bibnamefont {Heinzel}}, \
  and\ \bibinfo {author} {\bibfnamefont {T.}~\bibnamefont {Thurn-Albrecht}},\
  }\href@noop {} {\bibfield  {journal} {\bibinfo  {journal} {Appl. Phys.
  Lett.}\ }\textbf {\bibinfo {volume} {93}},\ \bibinfo {pages} {102110}
  (\bibinfo {year} {2008})}\BibitemShut {NoStop}%
\bibitem [{\citenamefont {Takahara}\ \emph {et~al.}(1995)\citenamefont
  {Takahara}, \citenamefont {Nomura}, \citenamefont {Gamo}, \citenamefont
  {Takaoka}, \citenamefont {Murase},\ and\ \citenamefont
  {Ahmed}}]{TakaharaJJAP95}%
  \BibitemOpen
  \bibfield  {author} {\bibinfo {author} {\bibfnamefont {J.}~\bibnamefont
  {Takahara}}, \bibinfo {author} {\bibfnamefont {A.}~\bibnamefont {Nomura}},
  \bibinfo {author} {\bibfnamefont {K.}~\bibnamefont {Gamo}}, \bibinfo {author}
  {\bibfnamefont {S.}~\bibnamefont {Takaoka}}, \bibinfo {author} {\bibfnamefont
  {K.}~\bibnamefont {Murase}}, \ and\ \bibinfo {author} {\bibfnamefont
  {H.}~\bibnamefont {Ahmed}},\ }\href@noop {} {\bibfield  {journal} {\bibinfo
  {journal} {Jap. J. of Appl. Phys.}\ }\textbf {\bibinfo {volume} {34}},\
  \bibinfo {pages} {4325} (\bibinfo {year} {1995})}\BibitemShut {NoStop}%
\bibitem [{\citenamefont {Geisler}\ \emph {et~al.}(2005)\citenamefont
  {Geisler}, \citenamefont {Chowdhury}, \citenamefont {Smet}, \citenamefont
  {H\"oppel}, \citenamefont {Umansky}, \citenamefont {Gerhardts},\ and\
  \citenamefont {von Klitzing}}]{GeislerPRB05}%
  \BibitemOpen
  \bibfield  {author} {\bibinfo {author} {\bibfnamefont {M.~C.}\ \bibnamefont
  {Geisler}}, \bibinfo {author} {\bibfnamefont {S.}~\bibnamefont {Chowdhury}},
  \bibinfo {author} {\bibfnamefont {J.~H.}\ \bibnamefont {Smet}}, \bibinfo
  {author} {\bibfnamefont {L.}~\bibnamefont {H\"oppel}}, \bibinfo {author}
  {\bibfnamefont {V.}~\bibnamefont {Umansky}}, \bibinfo {author} {\bibfnamefont
  {R.~R.}\ \bibnamefont {Gerhardts}}, \ and\ \bibinfo {author} {\bibfnamefont
  {K.}~\bibnamefont {von Klitzing}},\ }\href {\doibase
  10.1103/PhysRevB.72.045320} {\bibfield  {journal} {\bibinfo  {journal} {Phys.
  Rev. B}\ }\textbf {\bibinfo {volume} {72}},\ \bibinfo {pages} {045320}
  (\bibinfo {year} {2005})}\BibitemShut {NoStop}%

\bibitem [{\citenamefont {Heitmann}\ and\ \citenamefont
  {Kotthaus}(1993)}]{HeitmannPT93}%
  \BibitemOpen
  \bibfield  {author} {\bibinfo {author} {\bibfnamefont {D.}~\bibnamefont
  {Heitmann}}\ and\ \bibinfo {author} {\bibfnamefont {J.~P.}\ \bibnamefont
  {Kotthaus}},\ }\href {\doibase DOI:10.1063/1.881355} {\bibfield  {journal}
  {\bibinfo  {journal} {Physics Today}\ }\textbf {\bibinfo {volume} {46}},\
  \bibinfo {pages} {56} (\bibinfo {year} {1993})}\BibitemShut {NoStop}%

\bibitem [{\citenamefont {Chiu}\ \emph {et~al.}(1976)\citenamefont {Chiu},
  \citenamefont {Lee},\ and\ \citenamefont {Quinn}}]{ChiuSS76}%
  \BibitemOpen
  \bibfield  {author} {\bibinfo {author} {\bibfnamefont {K.}~\bibnamefont
  {Chiu}}, \bibinfo {author} {\bibfnamefont {T.}~\bibnamefont {Lee}}, \ and\
  \bibinfo {author} {\bibfnamefont {J.}~\bibnamefont {Quinn}},\ }\href
  {\doibase DOI: 10.1016/0039-6028(76)90132-1} {\bibfield  {journal} {\bibinfo
  {journal} {Surface Science}\ }\textbf {\bibinfo {volume} {58}},\ \bibinfo
  {pages} {182 } (\bibinfo {year} {1976})}\BibitemShut {NoStop}%

\bibitem{note3}
Photon absorption in a 2D superlattice is derived for example by 
V. Demikhovskii and A. Perov: J. Exp. Theor. Phys. {\bf 87}, 973 (1998), 
and a more general discussion can be found on p.~241 in
G. Bastard: Wave mechanics applied to semiconductor heterostructures, 
Wiley (1991).

\bibitem [{\citenamefont {Wang}\ \emph {et~al.}(2004)\citenamefont {Wang},
  \citenamefont {Vasilopoulos},\ and\ \citenamefont {Peeters}}]{WangPRB04}%
  \BibitemOpen
  \bibfield  {author} {\bibinfo {author} {\bibfnamefont {X.~F.}\ \bibnamefont
  {Wang}}, \bibinfo {author} {\bibfnamefont {P.}~\bibnamefont {Vasilopoulos}},
  \ and\ \bibinfo {author} {\bibfnamefont {F.~M.}\ \bibnamefont {Peeters}},\
  }\href {\doibase 10.1103/PhysRevB.69.035331} {\bibfield  {journal} {\bibinfo
  {journal} {Phys. Rev. B}\ }\textbf {\bibinfo {volume} {69}},\ \bibinfo
  {pages} {035331} (\bibinfo {year} {2004})}\BibitemShut {NoStop}%
\bibitem [{\citenamefont {Gradshteyn}\ and\ \citenamefont
  {Ryzhik}(1980)}]{Gradshteyn}%
  \BibitemOpen
  \bibfield  {author} {\bibinfo {author} {\bibfnamefont {I.~S.}\ \bibnamefont
  {Gradshteyn}}\ and\ \bibinfo {author} {\bibfnamefont {I.~M.}\ \bibnamefont
  {Ryzhik}},\ }\enquote {\bibinfo {title} {Table of integrals, series, and
  products},}\ \ (\bibinfo  {publisher} {Academic, New York},\ \bibinfo {year}
  {1980})\BibitemShut {NoStop}%
\bibitem [{\citenamefont {Stern}(1967)}]{SternPRL67}%
  \BibitemOpen
  \bibfield  {author} {\bibinfo {author} {\bibfnamefont {F.}~\bibnamefont
  {Stern}},\ }\href {\doibase 10.1103/PhysRevLett.18.546} {\bibfield  {journal}
  {\bibinfo  {journal} {Phys. Rev. Lett.}\ }\textbf {\bibinfo {volume} {18}},\
  \bibinfo {pages} {546} (\bibinfo {year} {1967})}\BibitemShut {NoStop}%
\bibitem [{\citenamefont {Mikhailov}\ and\ \citenamefont
  {Savostianova}(2005)}]{MikhailovPRB05}%
  \BibitemOpen
  \bibfield  {author} {\bibinfo {author} {\bibfnamefont {S.~A.}\ \bibnamefont
  {Mikhailov}}\ and\ \bibinfo {author} {\bibfnamefont {N.~A.}\ \bibnamefont
  {Savostianova}},\ }\href@noop {} {\bibfield  {journal} {\bibinfo  {journal}
  {Phys. Rev. B}\ }\textbf {\bibinfo {volume} {71}},\ \bibinfo {pages} {035320}
  (\bibinfo {year} {2005})}\BibitemShut {NoStop}%
\bibitem [{\citenamefont {Mikhailov}\ and\ \citenamefont
  {Savostianova}(2006)}]{MikhailovPRB06}%
  \BibitemOpen
  \bibfield  {author} {\bibinfo {author} {\bibfnamefont {S.~A.}\ \bibnamefont
  {Mikhailov}}\ and\ \bibinfo {author} {\bibfnamefont {N.~A.}\ \bibnamefont
  {Savostianova}},\ }\href {\doibase 10.1103/PhysRevB.74.045325} {\bibfield
  {journal} {\bibinfo  {journal} {Phys. Rev. B}\ }\textbf {\bibinfo {volume}
  {74}},\ \bibinfo {pages} {045325} (\bibinfo {year} {2006})}
\BibitemShut  {NoStop}%
\bibitem [{\citenamefont {Fedorych}\ \emph {et~al.}(2009)\citenamefont
  {Fedorych}, \citenamefont {Studenikin}, \citenamefont {Moreau}, \citenamefont
  {Potemski}, \citenamefont {Saku},\ and\ \citenamefont
  {Hirayama}}]{FedorychIJMPB09}%
  \BibitemOpen
  \bibfield  {author} {\bibinfo {author} {\bibfnamefont {O.~M.}\ \bibnamefont
  {Fedorych}}, \bibinfo {author} {\bibfnamefont {S.~A.}\ \bibnamefont
  {Studenikin}}, \bibinfo {author} {\bibfnamefont {S.}~\bibnamefont {Moreau}},
  \bibinfo {author} {\bibfnamefont {M.}~\bibnamefont {Potemski}}, \bibinfo
  {author} {\bibfnamefont {T.}~\bibnamefont {Saku}}, \ and\ \bibinfo {author}
  {\bibfnamefont {Y.}~\bibnamefont {Hirayama}},\ }\href@noop {} {\bibfield
  {journal} {\bibinfo  {journal} {Int. J. of Mod. Phys. B}\ }\textbf {\bibinfo
  {volume} {23}},\ \bibinfo {pages} {2698} (\bibinfo {year}
  {2009})}\BibitemShut {NoStop}%
\bibitem [{\citenamefont {Zhao}\ \emph {et~al.}(1992)\citenamefont {Zhao},
  \citenamefont {Tsui}, \citenamefont {Santos}, \citenamefont {Shayegan},
  \citenamefont {Ghanbari}, \citenamefont {Antoniadis},\ and\ \citenamefont
  {Smith}}]{ZhaoAPL92}%
  \BibitemOpen
  \bibfield  {author} {\bibinfo {author} {\bibfnamefont {Y.}~\bibnamefont
  {Zhao}}, \bibinfo {author} {\bibfnamefont {D.~C.}\ \bibnamefont {Tsui}},
  \bibinfo {author} {\bibfnamefont {M.}~\bibnamefont {Santos}}, \bibinfo
  {author} {\bibfnamefont {M.}~\bibnamefont {Shayegan}}, \bibinfo {author}
  {\bibfnamefont {R.~A.}\ \bibnamefont {Ghanbari}}, \bibinfo {author}
  {\bibfnamefont {D.~A.}\ \bibnamefont {Antoniadis}}, \ and\ \bibinfo {author}
  {\bibfnamefont {H.~I.}\ \bibnamefont {Smith}},\ }\href@noop {} {\bibfield
  {journal} {\bibinfo  {journal} {Appl. Phys. Lett.}\ }\textbf {\bibinfo
  {volume} {60}},\ \bibinfo {pages} {1510} (\bibinfo {year}
  {1992})}\BibitemShut {NoStop}%
\bibitem [{\citenamefont {Mikhailov}\ and\ \citenamefont
  {Volkov}(1995)}]{MikhailovPRB95}%
  \BibitemOpen
  \bibfield  {author} {\bibinfo {author} {\bibfnamefont {S.~A.}\ \bibnamefont
  {Mikhailov}}\ and\ \bibinfo {author} {\bibfnamefont {V.~A.}\ \bibnamefont
  {Volkov}},\ }\href@noop {} {\bibfield  {journal} {\bibinfo  {journal} {Phys.
  Rev. B}\ }\textbf {\bibinfo {volume} {52}},\ \bibinfo {pages} {17260}
  (\bibinfo {year} {1995})}\BibitemShut {NoStop}%
\bibitem{Yugova:2007_a}%
  \BibitemOpen
  \bibfield  {author} {\bibinfo {author} {\bibfnamefont {I. A.}\ \bibnamefont
  {Yugova}},\ \bibinfo {author} {\bibfnamefont {A.}\ \bibnamefont
  {Greilich}},\ \bibinfo {author} {\bibfnamefont {D. R.}\ \bibnamefont
  {Yakovlev}},\ \bibinfo {author} {\bibfnamefont {A. A.}\ \bibnamefont
  {Kiselev}},\ \bibinfo {author} {\bibfnamefont {M.}\ \bibnamefont
  {Bayer}},\ \bibinfo {author} {\bibfnamefont {V. V.}\ \bibnamefont
  {Petrov}},\ \bibinfo {author} {\bibfnamefont {Yu. K.}\ \bibnamefont
  {Dolgikh}},\ \bibinfo {author} {\bibfnamefont {D.}\ \bibnamefont
  {Reuter}}\ and \bibinfo {author} {\bibfnamefont {A. D.}\ \bibnamefont
  {Wieck}},\ }\href@noop {} {\bibfield  {journal} {\bibinfo  {journal} {Phys.
  Rev. B}\ }\textbf {\bibinfo {volume} {75}},\ \bibinfo {pages} {245302}
  (\bibinfo {year} {2007})}\BibitemShut {NoStop}%
\end{thebibliography}

%\end{document}

%

\end{document}